\documentclass[aps,pre,preprint]{revtex4-1}

\usepackage{amsmath}
\usepackage{graphicx}

\newcommand*{\defeq}{\mathrel{\rlap{%
                     \raisebox{0.3ex}{$\m@th\cdot$}}%
                     \raisebox{-0.3ex}{$\m@th\cdot$}}%
                     =}
\newcommand*{\defeqR}{=\mathrel{\rlap{%
                     \raisebox{0.3ex}{$\m@th\cdot$}}%
                     \raisebox{-0.3ex}{$\m@th\cdot$}}
                     }%

\newcommand{\ve}[1]{\boldsymbol{#1}}

\newcommand{\gaussd}[1]{\mathcal{D} #1}
\makeatletter
\renewcommand*{\defeq}{\mathrel{\rlap{%
                     \raisebox{0.3ex}{$\m@th\cdot$}}%
                     \raisebox{-0.3ex}{$\m@th\cdot$}}%
                     =}
\renewcommand*{\defeqR}{=\mathrel{\rlap{%
                     \raisebox{0.3ex}{$\m@th\cdot$}}%
                     \raisebox{-0.3ex}{$\m@th\cdot$}}%
                     }

\begin{document}


\title{How single neuron properties shape chaotic dynamics and \\ signal transmission in random neural networks}


\author{Samuel P. Muscinelli}
\email{samuel.muscinelli@epfl.ch}
\affiliation{School of Computer and Communication Sciences and School of Life Sciences \\
\'Ecole polytechnique f\'ed\'erale de 
Lausanne \\
Station 15, CH-1015 Lausanne EPFL, Switzerland}

\author{Wulfram Gerstner}
\affiliation{School of Computer and Communication Sciences and School of Life Sciences \\
\'Ecole polytechnique f\'ed\'erale de 
Lausanne \\
Station 15, CH-1015 Lausanne EPFL, Switzerland}

\author{Tilo Schwalger}
\affiliation{Bernstein Center for Computational 
Neuroscience, 10115 Berlin, Germany}
\affiliation{Institut f\"ur 
Mathematik, Technische Universit\"at Berlin, 10623 Berlin, Germany}

\begin{abstract}
While most models of randomly connected networks assume nodes with simple dynamics, nodes in realistic highly connected networks, such as neurons in the brain, exhibit intrinsic dynamics over multiple timescales.
We analyze how the dynamical properties of nodes (such as single neurons)  and recurrent connections interact to shape the effective dynamics in large randomly connected networks.
A novel dynamical mean-field theory for strongly connected networks of multi-dimensional rate units shows that the power spectrum of the network activity in the chaotic phase emerges from a nonlinear sharpening of the frequency response function of single units.
For the case of two-dimensional rate units with strong adaptation, we find that the network exhibits a state of ``resonant chaos'', characterized by robust, narrow-band stochastic oscillations.
The coherence of stochastic oscillations is maximal at the onset of chaos and their correlation time scales with the adaptation timescale of single units.
Surprisingly, the resonance frequency can be predicted from the properties of isolated units, even in the presence of heterogeneity in the adaptation parameters.
In the presence of these internally-generated chaotic fluctuations, the transmission of weak, low-frequency signals is strongly enhanced by adaptation, whereas signal transmission is not influenced by adaptation in the non-chaotic regime.
Our theoretical framework can be applied to other mechanisms at the level of single nodes, such as synaptic filtering, refractoriness or spike synchronization.
These results advance our understanding of the interaction between the dynamics of single units and recurrent connectivity, which is a fundamental step toward the description of biologically realistic network models in the brain, or, more generally, networks of other physical or man-made complex dynamical units.
\end{abstract}

\maketitle

\section{Introduction}\label{sec:introduction} 
Random network models of interacting dynamical elements (units, or nodes), are widely used across different scientific fields \cite{Mendes02}.
Examples from biology include neural network models \cite{Sompolinsky88}, metabolic networks \cite{Barkai97} and protein regulatory networks \cite{Kauffman93,Jeong01,Pomerance09}.
The dynamics of random networks also plays an important role in the study of epidemic outbreaks \cite{Pastor-Satorras01}, social networks \cite{Newman02}, power grids \cite{Nishikawa15} and transportation networks \cite{Banavar99}, as well as in abstract physics systems such as soft spin models \cite{Sompolinsky81} and networks of oscillators \cite{Rodrigues2016}.
In all the above examples, the collective dynamics result from the interplay between network connectivity and the dynamics of single units.

A typical property of large random networks of nonlinear dynamical elements is the existence of a chaotic phase.
In the context of neural networks, the rich dynamics of large random networks of neuron-like elements at the edge of chaos has been exploited to learn complex tasks involving generation of temporal patterns
\cite{Maass02a,Jaeger04,Sussillo09,Laje13,Nicola17,DePasquale18,Mastrogiuseppe18}.
In these and other related approaches, the chaotic behavior of the network mainly arises from the random interactions, whereas the dynamics of single elements are typically given by first-order differential equations.  
The simplicity of single elements allows to quantitatively determine the chaotic phase of the coupled elements using dynamical mean-field theory (DMFT) \cite{Sompolinsky88}, even in networks with more realistic connectivity structure \cite{Rajan06,Kadmon15,Mastrogiuseppe17,Mastrogiuseppe18}.

A fascinating question is what kind of activity emerges in brain-like networks, that are subject to additional biological constraints. 
Individual neurons exhibit rich multi-dimensional internal dynamics \cite{Benda03,Lundstrom08,La-Camera06,Pozzorini13} that are inconsistent with first-order equations.  
However, a theoretical understanding of the emergent activity patterns in networks of more realistic multi-dimensional dynamical elements is largely lacking.
In particular, beyond one-dimensional and simple oscillator models \cite{Sompolinsky88,WieBer15,vanMeegen18}, a self-consistent mean-field theory for fluctuations such as the autocorrelation function or the power spectrum of the network activity is still an unsolved theoretical problem. 
Here, we develop a theoretical framework that extends DMFT to multi-dimensional rate neurons.
Using this framework, we show that the power spectrum of the network activity in the nonlinear, strongly coupled regime, emerges from a sharpening of the single-neuron frequency response function due to strong recurrent connections.

Our theory uses firing rate models with two or more variables per unit.
While rate-based models \cite{Wilson72,Deco11} discard information on the exact spike-timing of single neurons, they have the advantage of being accessible to an analytical characterization of their dynamics.
However, commonly-used one-dimensional rate models cannot fully capture the dynamics of the mean activity of a population of spiking neurons, such as the synchronization of neurons in response to a stimulus onset \cite{Mainen95a,Bair96,DevRox17}, an effect that is readily observed in integrate-and-fire model \cite{Knight72,Konig96a,Gerstner00,Brette03,SchDeg17}.
To capture rapid synchronization after stimulus onset in rate models, it is necessary to consider at least two equations per rate neuron \cite{Mattia02,Schaffer13,Montbrio15}.
Multi-dimensional models also account for additional cellular mechanisms such as refractoriness \cite{BerMei98}, spike-frequency adaptation (SFA) \cite{Naud12,Schwalger13,DegSch14,SchDeg17}, synaptic filtering \cite{Fourcaud02,Schwalger08a}, subthreshold resonance \cite{Richardson03} or for the effect of dendritic compartments \cite{OstSza15,Doose16}.

To be specific, we focus on SFA, the decrease of a neuron's firing rate in response to a sustained stimulus, but our theory can also be applied to other phenomena.
SFA is present in neurons at all stages of sensory processing, and is believed to play a crucial role for efficient coding of external stimuli \cite{Benda03}.
Moreover, SFA over multiple timescales represents an efficient solution for information transmission of sensory signals whose statistics change dynamically \cite{Fairhall01,Lundstrom08,Pozzorini13}.
It is therefore of great interest to understand how adaptation and recurrent connections interact to shape network dynamics and signal transmission \cite{Mar99,Akerberg09}.
If connections and adaptation are weak, the network dynamics can be largely understood within linear response theory.
In particular, in the presence of signals and noise, linear response theory predicts that adaptation shapes signal and noise in precisely the same manner  \cite{DegSch14}, canceling the noise-shaping effect of adaptation \cite{Mar99,Akerberg09,Lindner16}.
In contrast, in strongly coupled networks generating chaotic fluctuations \cite{Sompolinsky88}, linear response theory is not applicable and the effect of adaptation on the signal transmission in this case remains poorly understood.
Here, we show that introducing adaptation into a strongly-coupled network of rate units shifts the network to a state of ``resonant'' chaos that is qualitatively different from the chaotic behavior of the network without adaptation.
In this state, the network generates a stable rhythm corresponding to a narrow-band peak in the power spectrum  which is robust against quenched disorder in adaptation parameters (heterogeneity).
We show that in this new regime the network has two interesting functional properties: first, the correlation time increases with the adaptation timescale; second, the low-frequency power of the chaotic activity is strongly decreased, enabling a better transmission of slow signals.

This paper is organized as follows: In section \ref{sec:microscopic} we present the microscopic network model and discuss its different dynamical regimes.
In sections \ref{sec:mean-field} and \ref{sec:chaotic-regime}, we present the mean-field theory and analyze the chaotic regime.
In each of these sections, we first present the general formalism and then apply the result to the case of adaptation with a single auxiliary variable.
Finally, we study the functional properties of the resonant chaotic state, focusing on the correlation time (section \ref{sec:corr_time}) and on the response to external~input~(section \ref{sec:driven}).
Detailed derivations and an example of a higher-dimensional rate model representing multi-timescale adaptation are provided in the appendix.

\section{Microscopic model}\label{sec:microscopic}

We are interested in studying the dynamics of a randomly connected recurrent network of multi-dimensional firing-rate units where each unit is described by a set of $D$ variables $x_i^1,\dotsc,x_i^D$.
The first variable $x_i^1$ is an activation variable that defines the output rate $y$ via a nonlinear function $\phi$, i.e. $y_i(t) = \phi(x_i^1(t))$.
More precisely, $\phi(x_i^1(t))$ should be interpreted as the deviation of the firing rate from some reference rate.
Therefore, $\phi(x_i^1(t))$ can take both positive and negative values. 

The remaining $D-1$ variables are auxiliary variables.
In isolation, each unit obeys a system of $D$ first-order linear differential equations
\begin{equation}\label{eq:micro-SU}
\dot{x}_i^\alpha(t) = \sum_{\beta=1}^{D}\mathrm{A^{\alpha\beta}}x_i^\beta(t) \quad ,
\end{equation}
where the dot denotes the temporal derivative.
In what follows, subscripts (in Latin letters) indicate the index of the unit in the network and run from 1 to $N$, while superscripts (in Greek letters) indicate the index of the variable in the rate model and run from 1 to $D$.
The matrix $\mathrm{A}$ is assumed to be non-singular 
and to have eigenvalues with negative real parts. 
We assume that the rate $\phi(x_j^1(t))$ is the only signal that unit $j$ uses to communicate with other units.
Conversely, the signals coming from other units only influence the variable $x_i^1$, i.e. the rate of unit $j$ is directly coupled only to the first variable of unit $i$.
The choice of having the same variable sending and receiving signals is dictated by simplicity and is not necessary for the development of the theory.
Unit $i$ receives input from all the other units, via a set of random connections $J_{ij}$, sampled i.i.d. from a Gaussian distribution with mean zero and variance $g^2/N$.
When incorporating these assumptions, the network equations read
\begin{equation}\label{eq:multi-d-micro}
\begin{aligned}
&\dot{x}_i^\alpha(t) = \sum_{\beta=1}^{D}\mathrm{A^{\alpha\beta}}x_i^\beta(t) + \delta^{\alpha1}\left(\sum_{j=1}^{N} J_{ij}\phi(x_j^1(t)) + I_i(t) \right)  \\
&J_{ij}\sim \mathcal{N}\left(0 , g^2/N \right)
\end{aligned}
\end{equation}
where $\delta^{\alpha\beta}$ is the Kronecker delta symbol. 
The external input $I_i(t)$ is assumed to have stationary statistics and zero mean.

\paragraph*{Network with adaptation.}
As a biologically relevant two-dimensional example, we consider rate units that undergo firing rate adaptation.  
We measure time in units of the timescale of the first variable such that $\mathrm{A}_{11}=-1$ (i.e. time is considered dimensionless), and work with a single auxiliary variable that mediates adaptation.
To ease our notation we drop the superscripts in the variables and write $x_i$ instead of $x_i^1$ and $a_i$ instead of $x_i^2$.
The adaptation variable $a_i(t)$ of neuron $i$ is driven by the neuron activation variable $x_i(t)$ and provides negative feedback onto $x_i$. 
The network equations (Eq.~\ref{eq:multi-d-micro}) for the adaptation case become
\begin{eqnarray}
\dot{x}_i(t) &=& -x_i(t) + \sum_{j=1}^N J_{ij} \phi(x_j(t)) - a_i(t) +I_i(t) \label{eq:micro-dynamics-x} \\
\dot{a}_i(t) &=& -\gamma a_i(t) + \gamma \beta x_i(t) \quad , \label{eq:micro-dynamics-a}
\end{eqnarray}
where the parameter $\gamma>0$ can be interpreted as the ratio of the timescales of the two variables $x$ and $a$, while $\beta>0$ is a parameter that controls the strength of adaptation.

Numerical simulations of the network with adaptation show that for low connection strength $g$, the network exhibits transient dynamics before it settles to a fixed point in which all $x_i$ are zero (Fig.~\ref{fig:microscopic}a,b).
By analyzing the stability of this fixed point (see appendix \ref{app:stability-microscopic} for details), we find that, in the $N\rightarrow\infty$ limit, the critical value of $g$ at which stability is lost depends on the adaptation parameters via
\begin{equation}\label{eq:g_critical}
g_c(\gamma,\beta) = 
\begin{cases}
\sqrt{1-\gamma(\gamma+2\beta) + 2\sqrt{\gamma^2\beta(2\gamma+2\beta+2)}},&\beta > \beta_H(\gamma) \\
1+\beta,&\beta\leq \beta_H(\gamma)
\end{cases}
\end{equation}
where $\beta_H(\gamma) = -1 -\gamma + \sqrt{2\gamma^2 + 2\gamma +1}$.
Notice that $g_c(\gamma,\beta)>1$ for all $\gamma,\beta >0$, i.e. adaptation stabilizes the dynamics since in the case without adaptation we have $g_c(0,0)=1$ \cite{Sompolinsky88}.
Interestingly, the two different cases in Eq.~\ref{eq:g_critical} correspond to two different bifurcation types: for $\beta<\beta_H(\gamma)$ we find that at $g=g_c(\gamma,\beta)$ we have a saddle-node bifurcation.
In contrast, if $\beta>\beta_H(\gamma)$, the system undergoes a Hopf bifurcation (see appendix \ref{app:stability-microscopic}), whose characteristic frequency is given by
\begin{equation}\label{eq:res_freq_micro_maintext}
f_m = \frac{1}{2\pi}\sqrt{-\gamma^2 +\sqrt{\beta\gamma^2(\beta+2\gamma+2)}} \quad .
\end{equation}
Examples of eigenvalue spectra of the randomly coupled network close to these two different bifurcations are shown in the insets of Fig.~\ref{fig:microscopic}:
For $\beta>\beta_H(\gamma)$, the eigenvalue spectrum is deformed such that the eigenvalues with the largest real part are complex (Fig.~\ref{fig:microscopic}a,c).
In contrast, for $\beta<\beta_H(\gamma)$ the spectrum is convex, exhibiting a rightmost eigenvalue that is real (Fig.~\ref{fig:microscopic}b,d). 
Above the bifurcation, i.e. for $g>g_c(\gamma,\beta)$, the network exhibits self-sustained, irregular fluctuations (Fig.~\ref{fig:microscopic}c,d) that we will characterize in the next sections.

In all the simulations and numerical integrations, we choose $\phi(x)$ as a piecewise-linear function given by
\begin{equation}\label{eq:PWL_def}
\phi_{PL}(x) = \begin{cases}
-1 \quad \text{for} \quad x<-1 \\
x \quad \text{for} \quad -1<x<1 \\
1 \quad \text{for} \quad x>1
\end{cases}
\end{equation}
unless stated otherwise.

\begin{figure*}[!ht]
\begin{center}
\includegraphics[scale=1]{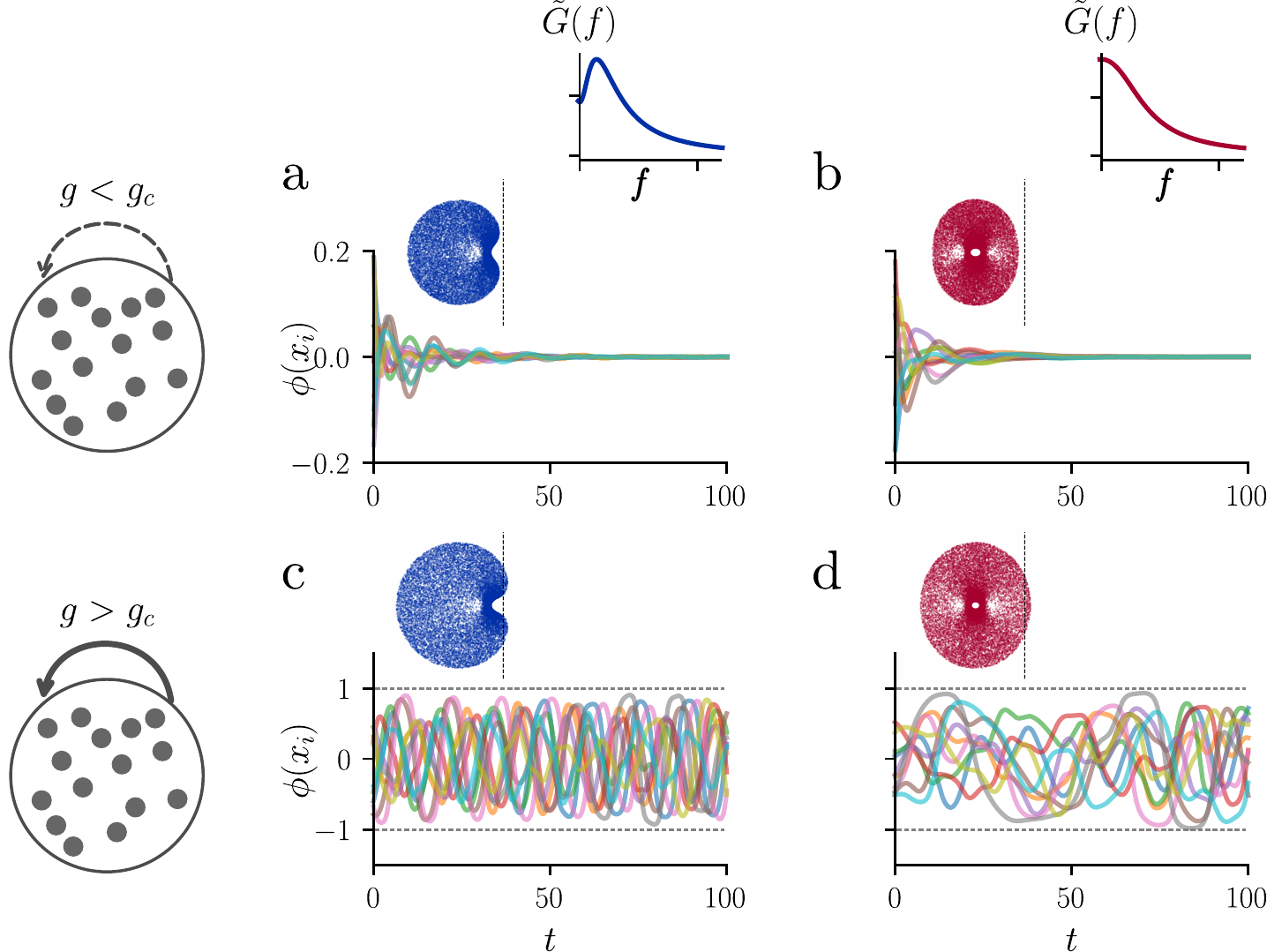}
\caption[Microscopic network dynamics with single-scale adaptation]{\textbf{Microscopic network dynamics with single-scale adaptation.}
In the top row (panels \textbf{a} and \textbf{b}), the network is below the bifurcation, ($g=0.96g_c(\gamma,\beta)$), and it exhibits a transient activity to the stable fixed point.
In the bottom row (panels \textbf{c} and \textbf{d}), the fixed point is unstable ($g=1.3g_c$) and the network exhibits irregular, self-sustained oscillations. 
In the left column (panels \textbf{a} and \textbf{c}, the network is in the resonant regime ($\gamma=0.2$, $\beta=0.5$), as it can be seen from the single-neuron linear frequency response function $\tilde{G}(f)$ (cf. Eq.~\ref{eq:chi0_def_adapt}).
In the right column (panels \textbf{b} and \textbf{d}), the network is in the non-resonant regime ($\gamma=1$, $\beta=0.1$).
For each panel, ten randomly chosen units are shown, out of $N=1000$ units.
Panel \textbf{c} corresponds to the resonant chaotic state, while in panel \textbf{d} the system exhibits chaotic activity similar to the case described in \cite{Sompolinsky88}.
The insets show the eigenvalue spectrum in the complex plane for the four different sets of parameters.
The dashed black line indicates the imaginary axis.
Comparing the eigenvalue spectrum of panel \textbf{a} with the one of panel \textbf{c}, we see that the network undergoes a Hopf bifurcation.  
}
\label{fig:microscopic}
\end{center}
\end{figure*}

\section{Mean-field theory}{\label{sec:mean-field}}

\paragraph*{General theory.}
The dynamics of the $ND$-dimensional dynamical system in Eq.~\ref{eq:multi-d-micro} for large $N$ is too high-dimensional to be studied at the microscopic level.
In contrast, using dynamical mean-field theory \cite{Sompolinsky88}, we can find properties of the network dynamics that are independent of the specific connectivity realization.
In what follows, we will assume that the external input $I_i(t)$ to each unit is an independent realization of the same Gaussian process.
Following \cite{Sompolinsky88}, we approximate the network input to a representative unit $i$ with a Gaussian process $\eta$ and substitute the average over time, initial conditions and network realizations with the average over realizations of $\eta$.
This approximation is valid in the large-$N$ limit, in which neurons become independent \cite{Schucker16a,Crisanti18}.
In the mean-field description, the activity of each individual unit in the network follows a realization of the following system of $D$ stochastic differential equations, to which we refer to as mean-field equations (see appendix \ref{app:DMFT_derivation} for more details)
\begin{equation}\label{eq:multi-d-mean-field}
\dot{x}^\alpha(t) = \sum_{\beta=1}^D \mathrm{A}^{\alpha\beta} x^\beta(t)  + \delta^{\alpha 1}\left( \eta(t) +I(t)\right) \quad , 
\end{equation}
where $\eta(t)$ is a Gaussian process.
The mean $\langle\eta(t)\rangle$ vanishes because the averaging over the Gaussian process statistics mimics the average over different neurons and network realizations, and the connections $J_{ij}$ in Eq.~\ref{eq:multi-d-micro} are sampled from a Gaussian distribution with mean zero.
Thanks to the fact that neurons become independent in the large-$N$ limit \cite{Sompolinsky88}, the average of the network input over network realizations is also zero (see appendix~\ref{app:DMFT_derivation} for more details). 
On the other hand, the autocorrelation function $\langle \eta(t)\eta(s) \rangle$ needs to be determined self-consistently by imposing (cf. appendix \ref{app:DMFT_derivation})
\begin{equation}\label{eq:multi-d-self-cons}
\langle \eta(t)\eta(s) \rangle = g^2 \langle \phi(x^1(t))\phi(x^1(s))\rangle \quad .
\end{equation}
Thanks to the mean-field approximations, we reduced the $ND$-dimensional, deterministic, nonlinear system of Eq.~\ref{eq:multi-d-micro} to the D-dimensional, stochastic system of Eq.~\ref{eq:multi-d-mean-field}, which looks linear at first glance.
However, the nonlinearity is important and is hidden in the self-consistent match of the second moment, as expressed by Eq.~\ref{eq:multi-d-self-cons}.
The linear mathematical structure of Eq.~\ref{eq:multi-d-mean-field} allows us to write, in the frequency domain
\begin{equation}\label{eq:mean-field-essential-GM}
\tilde{x}^1(f) = \tilde{\chi}_0(f) \left( \tilde{\eta}(f) + \tilde{I}(f)\right) \quad,
\end{equation}
where $\tilde{\chi}_0(f)$ is the linear response function (susceptibility) of the mean-field system (Eq. \ref{eq:multi-d-mean-field}), which is equal to the linear response function of an uncoupled single neuron in the microscopic description (Eq.~\ref{eq:micro-SU}).
For the linear dynamics given by Eq.~\ref{eq:multi-d-mean-field}, the linear response function $\tilde{\chi}_0(f)$ is given by
\begin{equation}
\tilde{\chi}_0(f) = \left[\left(2\pi i f \mathrm{I}_D - \mathrm{A} \right)^{-1} \right]^{1,1} \quad ,
\end{equation}
where $\mathrm{I}_D$ is the $D$-dimensional identity matrix and the upper indices 1,1 indicate the first element of the first row of the matrix inside the square brackets.

In what follows, we assume that the external input $I(t)$ is stationary and zero-mean, and that the network is in the stationary regime.
Therefore, the mean of all variables is equal to zero. 
The second-order statistics must be determined self-consistently.
In the frequency domain, this requires a self-consistent determination of the power spectral density (``power spectrum'' for short) $S_x(f)$ of the activation variable $x^1$, defined as the Fourier transform of the autocorrelation function, $S_x(f) = \int_{-\infty}^\infty e^{-2\pi if\tau}\langle x^1(t+\tau)x^1(t)\rangle\,d\tau$. 
Using the squared modulus of the linear response function $\tilde{G}(f) \defeq |\tilde{\chi}_0(f)|^2$, the power spectrum can be expressed as
\begin{equation}\label{eq:spectrum-matrix-form}
S_x(f) = \tilde{G}(f) \left(S_\eta(f) + S_I(f)\right),
\end{equation}
where $S_\eta(f)$ and $S_I(f)$ denote the power spectral densities of $\eta(t)$ and $I(t)$, respectively.
Importantly, from Eq.~\ref{eq:multi-d-self-cons} we have that $S_\eta(f)$ depends implicitly on $S_x(f)$ through the self-consistency condition
\begin{equation}\label{eq:multi-d_selfcons_fourier}
S_\eta(f) = g^2 S_{\phi(x^1)}(f).
\end{equation}
The factor $\tilde{G}(f)$ can be expressed as a function of the matrix $\mathrm{A}$ as
\begin{equation}
\tilde{G}(f)=\frac{\left|\left[\text{adj}(2\pi i f \mathrm{I}_D -\mathrm{A})\right]^{1,   1}\right|^2}
{\prod_{i=1}^D|2\pi i f-\lambda_\mathrm{A}^i|^2},
\end{equation}
where $\text{adj}\left(2\pi i f\mathrm{I}_D - \mathrm{A} \right)$ is the adjoint matrix of $\left(2\pi i f\mathrm{I}_D - \mathrm{A} \right)$ and $\lambda_\mathrm{A}^i$ are the eigenvalues of $\mathrm{A}$.
In appendix \ref{app:DMFT_FP_stability}, we show that knowing the maximum of  $\tilde{G}(f)$ is sufficient to compute the critical value of the coupling $g_c$:
\begin{equation}
g_c^2 \max_f \tilde{G}(f) = 1 \quad .
\end{equation}

\paragraph*{Network with adaptation.}
For the network with adaptation defined by Eqs. (\ref{eq:micro-dynamics-x}, \ref{eq:micro-dynamics-a}), the mean-field equations read
\begin{align}
\dot{x}(t) =& -x(t) -a(t) + \eta(t) +I(t) \label{eq:mean-field-x}\\
\dot{a}(t) =& -\gamma a(t) + \gamma\beta x(t) \quad , \label{eq:mean-field-a}
\end{align}    
with $\langle \eta(t) \rangle = 0$ and $\langle \eta(t+\tau)\eta(t)\rangle = g^2 \langle \phi(x(t+\tau))\phi(x(t))\rangle$.
The self-consistent equations (Eqs. (\ref{eq:spectrum-matrix-form}, \ref{eq:multi-d_selfcons_fourier}) ), reduce to
\begin{equation}\label{eq:spectrum_self_consistent-A}
S_x(f) = \tilde{G}(f) \left( g^2 S_{\phi(x)}(f) + S_I(f) \right)\quad.
\end{equation}
The factor $\tilde{G}(f)$ can be calculated explicitly, yielding
\begin{equation}\label{eq:chi0_def_adapt}
\tilde{G}(f) = \frac{\gamma^2 + \omega^2}{\omega^4 + (1+\gamma^2 - 2\beta\gamma)\omega^2 + \gamma^2(1+\beta)^2} \quad ,
\end{equation}
with $\omega=2\pi f$.

In the next section, we show how the qualitative features of the dynamics of the network in the fluctuating regime can be predicted by the properties of the single unit linear response function, as summarized in the factor $\tilde{G}(f)$.

\section{Resonant chaos in networks with adaptation}\label{sec:chaotic-regime}

The traditional approach in the DMFT literature is to consider the time-domain version of Eq.~\ref{eq:spectrum-matrix-form} \cite{Sompolinsky88}.
Applying the inverse Fourier transform to Eq.~\ref{eq:spectrum-matrix-form} would lead to a differential equation of order $2D$. 
Unfortunately, by contrast with the case $D=1$, for the multi-dimensional case $D>1$ the dynamics is no longer conservative, which precludes the determination of the initial conditions (see \cite{Sompolinsky88}).
We propose an alternative approach to find a self-consistent solution to Eq.~\ref{eq:spectrum-matrix-form} in the Fourier domain.
This approach is based on an iterative map, the fixed point of which is the self-consistent solution.  
Iterative methods have been proposed previously both in the context of spiking \cite{DumWie14,WieBer15} and rate-based networks \cite{Stern14} using Monte-Carlo methods.
Here, we use a semi-analytical iteration method that allows to rapidly solve for the self-consistent power spectrum, and hence to qualitatively understand several features of the network dynamics.

\begin{figure*}[!ht]
\includegraphics[width=1.0\textwidth]{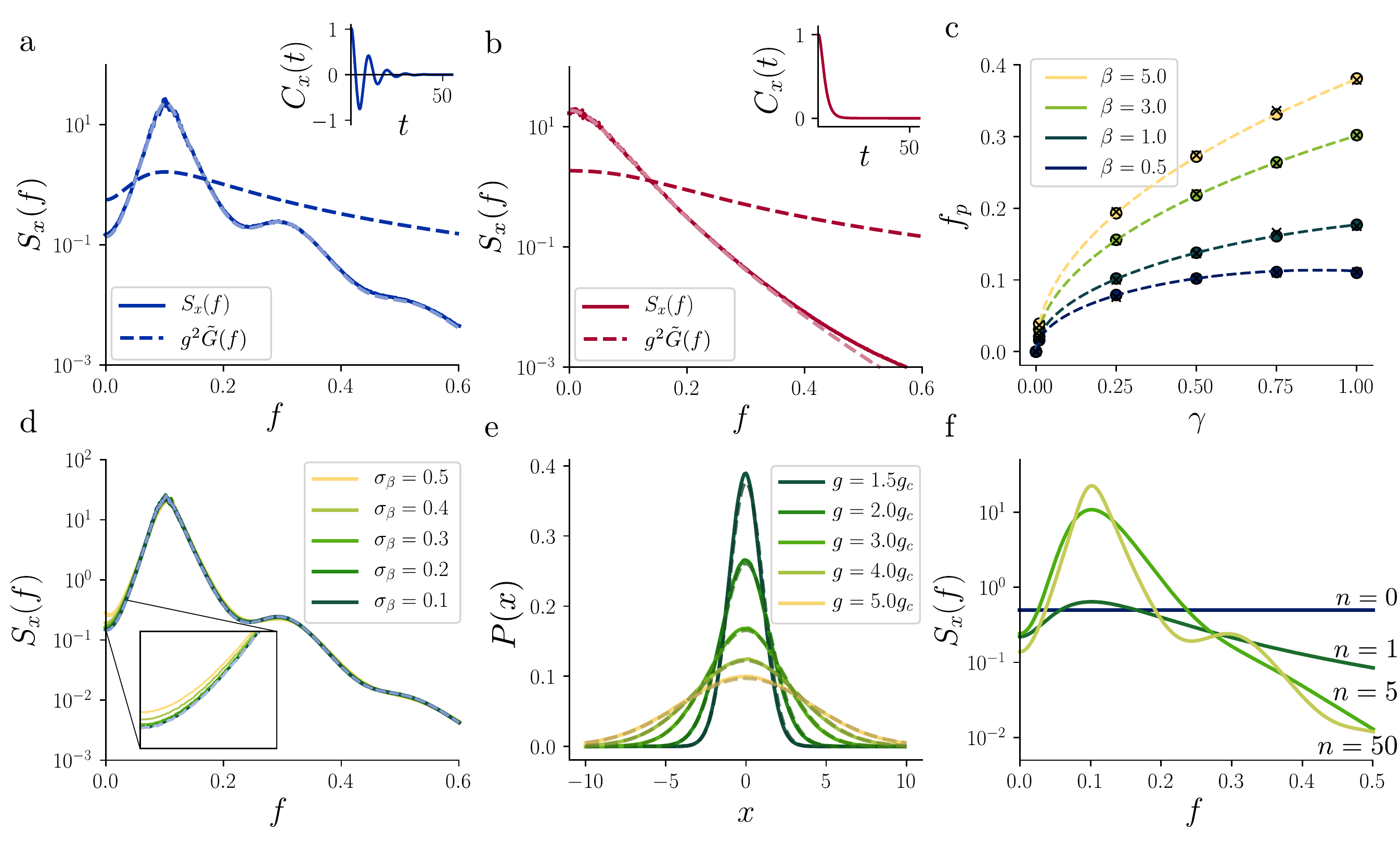}
\caption[Dynamical regimes in the mean-field description]{\textbf{Self-consistent statistics in the chaotic regime.}
\textbf{a:} Resonant (narrow-band) chaos.
Power spectral density obtained from mean-field theory (solid line) and microscopic simulations (light blue, dashed) for  $\gamma=0.25$, $\beta=1$ and $g = 2 g_c(\gamma, \beta)$.
The dashed, dark blue line indicates the square modulus of the linear response function $\tilde{G}(f)$ for the same adaptation parameters.
Inset: Normalized mean-field autocorrelation $C_x(\tau)$ for the same parameters, plotted against the time lag in units of $\tau_x$.
\textbf{b:} Non-resonant (broad-band) chaotic regime.
Curves and inset are the same as in \textbf{a}, but with $\gamma=1$, $\beta=0.1$ and $g=2 g_c(\gamma, \beta$).
\textbf{c:} Maximum-power frequency $f_p$ of the recurrent network plotted against $\gamma$, for different $\beta$.
Crosses depict results obtained from microscopic simulations, circles show the semi-analytical prediction based on the iterative method and dashed lines shows the theory based on the single neuron response function.
For $\gamma=0$ all curves start at $f_p=0$.
\textbf{d:} Power spectral density $S_x(f)$ for different levels of heterogeneity of the parameter $\beta$ (solid lines), compared to the case without heterogeneity (dashed line).
All the curves are almost superimposed, except at very low frequencies where small deviations are visible (inset).
Parameters: $\gamma=0.25$, $\bar{\beta}=1$, $g=2g_c(\gamma,\bar{\beta})$.
\textbf{e:} Distributions $P(x)$ of the activation $x$ from microscopic simulation ($N=2000$, solid lines) and theoretical prediction (dashed lines).
The adaptation parameter were $\gamma=0.25$ and $\beta=1$.
\textit{Caption continues on the next page.}
}
\label{fig:mean_field_regimes}
\end{figure*}

In the frequency domain, the linear transform associated with $\tilde{G}(f)$ is simple, whereas the nonlinearity $\phi(x)$ is difficult to handle.
Concretely, we need to express $S_{\phi(x^1)}$ as a functional of $S_x(f)$.
This calculation can be performed semi-analytically for the piecewise-linear nonlinearity (a detailed treatment of the nonlinear step is given in appendix~\ref{app:nonlinearities}).
The idea of our iterative method is to start with an arbitrary initial power spectral density $S_{\phi(x^1)}^{(0)}(f)$, which we choose to be constant (white noise).
We then apply multiple iterations each consisting of a linear step followed by a nonlinear one (Fig.~\ref{fig:mean_field_regimes}f).
At each iteration, the linear step is simply a multiplication by $g^2\tilde{G}(f)$ and it allows us to compute $(S_x)^{(n+1)}(f)$.
The nonlinear step afterwards transforms $(S_x)^{(n+1)}(f)$ into $S_{\phi(x^1)}^{(n+1)}(f)$.

By studying the iterative map that defines the mean-field solution, we conclude that the power spectrum of the network activity emerges from a sharpening of the linear response function $\tilde{G}(f)$ of single units.
The sharpening mainly arises from repeated multiplications with the factor $g^2\tilde{G}(f)$ in the iteration, which however is balanced by cross-frequency interactions and saturation effects of the nonlinear steps (see appendix \ref{app:qualitative} for a detailed discussion).
As a result, the network activity exhibits the same frequency bands that are preferred by single neurons, albeit much narrower.

\addtocounter{figure}{-1}
\begin{figure*}[ht!]
\caption{\textit{Continues from previous page.}
\textbf{f:} Power spectral density $S_x$ at different iterations $n$, for the network with adaptation with the same parameters as in \textbf{a}.
The initial power spectral density is a constant.
At $n=50$ the iterative map has converged.}
\noindent\makebox[\linewidth]{\rule{\textwidth}{0.4pt}}
\end{figure*}

We apply the iterative method to solve the mean-field equations for the network with adaptation (Eq.~\ref{eq:spectrum_self_consistent-A}), in the absence of external input ($I=0$).
We find that if $g<g_c(\gamma,\beta)$, the power spectrum converges to zero, $S_x(f)\rightarrow 0$, at all frequencies.
Therefore the mean-field variable $x$ is constantly equal to zero.
This is consistent with the presence of a stable fixed point at zero and it indicates that, in the thermodynamic limit, the fixed point solution is the only possible one.
In this regime we can calculate the mean-field linear response function $\tilde{\chi}(f)$ of the network, which is given by (see appendix \ref{app:DMFT_FP_stability})
\begin{equation}
\tilde{\chi}(f) = \frac{\tilde{G}(f)}{1-g^2 \tilde{G}(f)} \quad .
\end{equation}
i.e. the network has a sharper linear response function than the single units.

On the other hand, if $g>g_c(\gamma,\beta)$, the mean-field network is characterized by a nonzero, continuous power spectral density (Fig.~\ref{fig:mean_field_regimes}).
This is an indication that, at the microscopic level, the network is in a chaotic state \cite{Schuecker18}.
However, we stress that a more rigorous proof of chaos would require the computation of the maximum Lyapunov exponent of the network, which we will not perform.
In contrast to a network without adaptation \cite{Sompolinsky88}, we find that in the presence of adaptation the network can be in two qualitatively different chaotic regimes.
For very weak and/or fast adaptation, the chaotic fluctuations are qualitatively the same as for the network without adaptation, i.e. the power spectrum is broad-band with maximum at $f=0$ (Fig. \ref{fig:mean_field_regimes}b).
We refer to this regime as to the non-resonant regime.
On the other hand, for strong and/or slow adaptation, the mean-field network settles in a new regime, 
characterized by an autocorrelation that decays to zero via damped oscillations and, equivalently, by a power spectrum that exhibits a pronounced resonance band around a nonzero resonance frequency $f_p$ (Fig. \ref{fig:mean_field_regimes}a).
The decaying autocorrelation function and the continuous power spectral density are an indication that the network is --~also in this regime -- in a state of microscopic chaos.
This new dynamical state, that we refer to as \textit{resonant chaos}, is qualitatively different from the one of the non-resonant regime and from the one of the non-adaptive network.

Strikingly, whether the network settles in the resonant or in the non-resonant regime can be predicted purely based on the single-unit adaptation properties.
More precisely, if $\beta<\beta_H(\gamma)$, the function $\tilde{G}(f)$ is monotonically decreasing with the frequency $f$, i.e. it exhibits a low-pass characteristic  (Fig. \ref{fig:mean_field_regimes}b).
This low-pass behavior of the single neuron is reflected by a power spectrum of the network that is also dominated by low frequencies, albeit less broad.
The network power spectrum corresponds exactly to the non-resonant regime discussed above.

In contrast, if $\beta>\beta_H(\gamma)$, the single neuron response amplitude $\tilde{G}(f)$ exhibits a maximum at a nonzero frequency $f_0 = \frac{1}{2\pi}\sqrt{-\gamma^2 + \sqrt{\beta\gamma^2(\beta+2\gamma+2)}}$. 
Such a resonance peak is typical of a band-pass filter (Fig. \ref{fig:mean_field_regimes}a).
The frequency $f_0$ is identical to $f_m$ (Eq.~\ref{eq:res_freq_micro_maintext}), which is derived from the imaginary part of the critical eigenvalue at the Hopf bifurcation (see appendix \ref{app:stability-microscopic}).
The single-neuron linear response characteristics are qualitatively preserved in the fluctuating activity of the recurrent network, which also exhibits a power spectral density dominated by a nonzero frequency $f_p$.
This regime corresponds to the resonant regime discussed above.
Interestingly, we find numerically that $f_p=f_0$, i.e. the resonance frequency is not affected by the introduction of recurrent connections  (Fig. \ref{fig:mean_field_regimes}c, tested up to $g=5g_c(\gamma,\beta)$).
We notice that the non-resonant and resonant regimes are consistent with the fixed point stability analysis of the network in the microscopic description. 
Indeed, the resonant and non-resonant regimes match the regions in which we observe Hopf or saddle-node bifurcations, respectively (appendix \ref{app:stability-microscopic}).

Using simulations of the full microscopic network, we verify that the mean-field description is a good approximation of the system for large but finite $N$.
In Fig. \ref{fig:mean_field_regimes}e we show that the probability density of the activation variable $x$ measured from the microscopic simulations matches the Gaussian distribution predicted by the mean-field theory, with relatively small finite-size effects that increase close to the criticality (see Fig.~\ref{fig:mean_field_regimes}e, $g=1.5g_c(\gamma,\beta)$).
Moreover, the mean-field solution provides a good description of the system for a wide range of adaptation parameters $\gamma\, , \, \beta$ (Fig. \ref{fig:mean_field_regimes}c).

\paragraph*{Network with heterogeneous adaptation.}
The narrow-band oscillations in networks of adapting neurons reported so far have been obtained for networks of identical neurons.
The variability of physiological properties of real neurons, however, suggests that adaptation parameters differ among neurons.
Heterogeneity of neuronal parameters is known to sensitively influence synchronization in networks of neural oscillators \cite{Golomb01a}.
In particular, mismatches of oscillation frequencies can impede the formation of neural rhythms.
Does heterogeneity have a similar effect in strongly coupled random networks of adapting neurons?
To address this question, we introduce a second source of disorder in the system by considering quenched randomness in the adaptation parameters.
Specifically, we construct the heterogeneous network by adding Gaussian noise to the parameter $\beta$, i.e. by sampling $\beta\sim \mathcal{N}\left(\bar{\beta},\sigma_\beta^2\right)$ independently for each neuron.
Numerical simulations of the network in the presence of heterogeneous adaptation show that the dynamics of the random network are surprisingly robust to this type of noise (Fig. \ref{fig:mean_field_regimes}d).
Even for relatively high variability ($\sigma_\beta/\bar{\beta}$=0.5), the only effect is a barely visible increase of the power spectral density at low-frequencies (Fig. \ref{fig:mean_field_regimes}d, inset).
In appendix~\ref{app:heterogeneous} we derive the mean-field equations that correspond to the network with heterogeneous adaptation, and we compute the effective factor $\tilde{G}_H(f)$ in this case.
The semi-analytical solution of the mean-field theory for heterogeneous adaptation predicts, similar to simulations, a stronger power at low frequencies than in the homogeneous case.
However, the deviations predicted by the theory are smaller than the mismatch between theory and simulations for the homogeneous case, so that we could not perform a quantitative verification of the mean-field theory for the heterogeneous case.

We now focus on the resonant chaotic regime, that represents the novel dynamical state that emerges from the introduction of adaptation and study the functional properties of the network in this regime.

\section{Correlation time and coherence of the oscillations}\label{sec:corr_time}

\begin{figure*}[!ht]
\begin{center}
\includegraphics[width=1\textwidth]{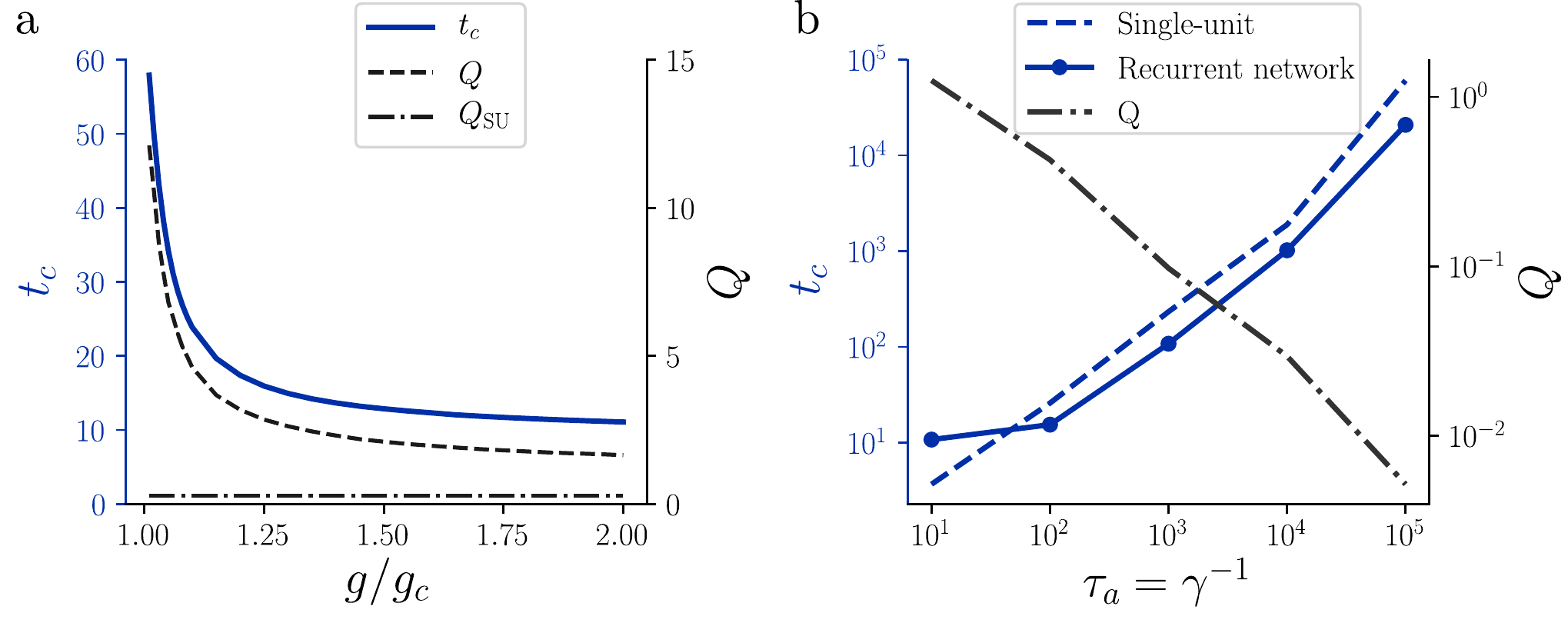}
\caption[Correlation time and effect of recurrent connections]{\textbf{Correlation time and effect of recurrent connections.}
\textbf{a:} Correlation time (blue solid line) and Q-factor (dashed line) as a function of the connectivity strength.
The weakest connectivity level plotted is $g=1.1g_c(\gamma,\beta)$.
Adaptation parameters: $\gamma=0.1$ and $\beta=1$. 
The dash-dotted horizontal line indicates the Q-factor of a single unit with the same adaptation parameters, driven by white noise.
\textbf{b:} Correlation time (blue) and Q-factor (black, dash-dotted line) as a function of the adaptation timescale $\tau_a \defeq \gamma^{-1}$.
Both the recurrent network (solid line) and the single unit driven by white noise (dashed line) scale with $\tau_a$.
$\beta=1$ and $g = 1.5g_c(\gamma,\beta)$.
}
\label{fig:corr_time}
\end{center}
\end{figure*}

While the resonance frequency in the resonant regime seems to depend solely on the single-neuron properties, the introduction of recurrent connections increases the coherence of the stochastic oscillations, i.e. decreases the width of the resonance band.
The narrower the resonance band, the more coherent the oscillatory behavior will be.
To quantify the increase of the oscillation coherence, we measure the  quality factor (Q-factor) of the stochastic oscillations, defined as
\begin{equation}
Q = \frac{f_p}{\Delta f_{\text{HM}}} \quad ,
\end{equation}
where $\Delta f_{\text{HM}}$ is the frequency width of the power spectrum $S_x(f)$ at the half-maximum. 
Intuitively, for a narrow-band oscillation, the quality factor quantifies the number of oscillation cycles during the characteristic decay time of the autocorrelation function.
For a single neuron driven by {\em white noise} ($\langle \eta(t)\eta(t')\rangle=\delta(t-t')$), the single-neuron power spectrum of $x$ is proportional to $\tilde{G}(f)$. 
Compared to this reference shape, we find a higher Q-factor in the recurrent network (Fig. \ref{fig:corr_time}a), corresponding to a sharper resonance peak in the power spectrum (see also appendix \ref{app:qualitative}).
When approaching the criticality from the chaotic phase, $g\rightarrow g_c(\gamma,\beta)^+$, the quality factor diverges (Fig. \ref{fig:corr_time}a), i.e. the dynamics approach regular oscillations. 

While the Q-factor measures the decay time constant of the autocorrelation function relative to the mean oscillation period, it is also interesting to consider the absolute correlation time of the activity. 
As a measure of correlation time of a stochastic process we use the normalized first moment (center of mass) of the absolute value of the autocorrelation function (e.g. \cite{LinWes89}),
\begin{equation}
t_c = \frac{\int_0^\infty \tau \left|C_x(\tau)\right|d\tau}{\int_0^\infty \left|C_x(\tau)\right|d\tau} \quad .
\end{equation}
Since the Q-factor diverges when $g\rightarrow g_c(\gamma,\beta)$, in this limit the corresponding autocorrelation exhibits sustained oscillations with a diverging correlation time. 
Due to the increase of the Q-factor, the correlation time also diverges when $g\rightarrow g_c(\gamma,\beta)$ (Fig. \ref{fig:corr_time}a). 

In the regime of slow adaptation, a single unit driven by white noise can have a larger correlation time than a recurrent network (Fig. \ref{fig:corr_time}b).
This is due to the fact that in this regime the correlation time of the single unit driven by white noise is dominated by the long tail of the autocorrelation.
The introduction of recurrent connections increases the oscillatory component, giving a larger ``weight'' to the short time lags, thus decreasing $t_c$.
Nevertheless, the correlation time increases with the timescale of adaptation $\tau_a$ for both the single unit driven by white noise and the recurrent network (Fig. \ref{fig:corr_time}b).
Note that the Q-factor goes to zero for very large adaptation timescale ($\gamma\rightarrow 0$), so that the dominant contribution to the correlation time in this regime is the non-oscillatory one.

\section{Response to periodic stimuli}\label{sec:driven}

\begin{figure*}[!ht]
\begin{center}
\includegraphics[width=0.99\textwidth]{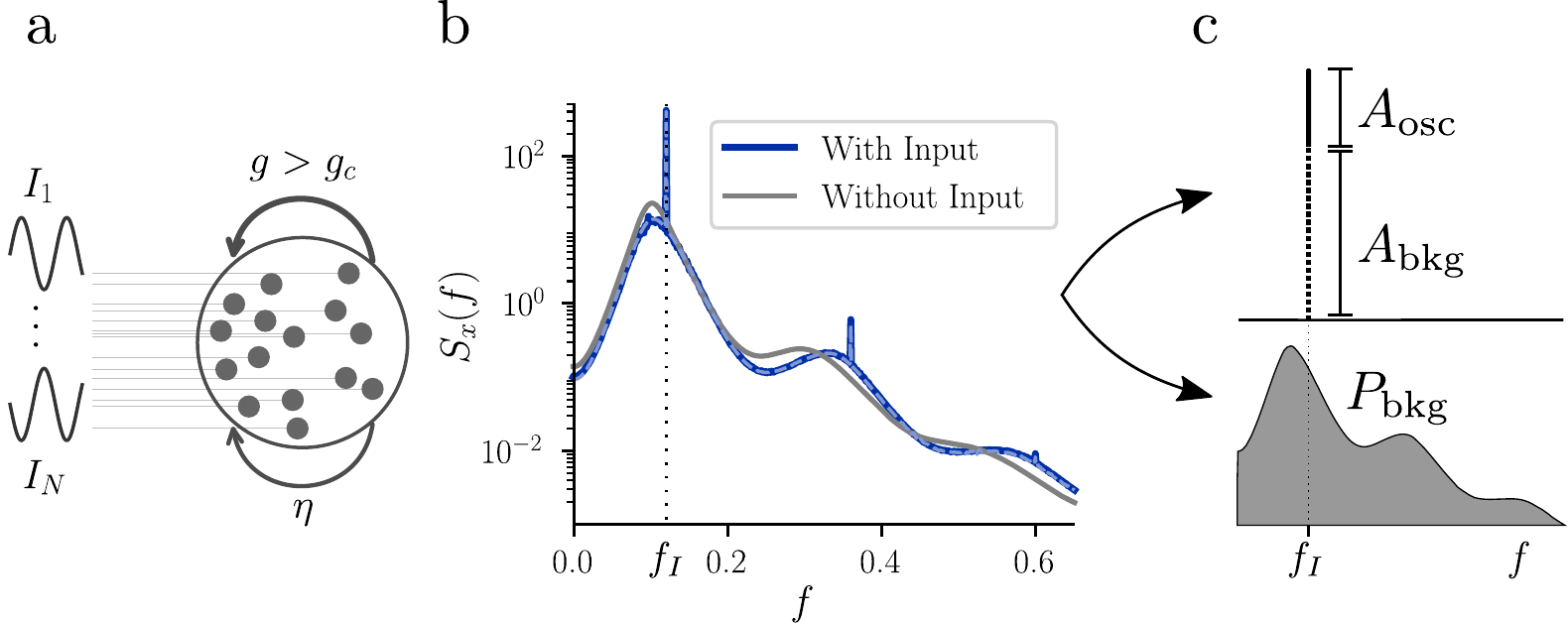}
\end{center}
\caption[Response of the mean-field network to an oscillatory input]{\textbf{Response of the mean-field network to an oscillatory input.}
\textbf{a:} Schematic representation of the random network driven by an external input, with phase randomization.
For $g>g_c$, the chaotic activity can be seen as internally-generated noise.
\textbf{b:} Effect of an oscillatory external input on the power spectral density $S_x(f)$.
In the example, $\gamma=0.25$, $\beta=1$, $g=2g_c(\gamma,\beta)$, $f_I = 0.12$, while $A_I=0.5$ (blue) and $A_I=0$ (gray).
Simulations (solid blue) and theory (dashed blue) are superimposed.
\textbf{c:} Top: Schematic representation of the separation of the power spectral density into its oscillatory ($A_\text{osc}$) and chaotic ($A_\text{bkg}$) components.
Note that these quantities depend on the size of the frequency discretization bin.
Bottom: Graphical interpretation of $P_\text{bkg}$, i.e. the total variance of the network activity due to chaotic activity (shaded gray area).
}
\label{fig:driven_net}
\end{figure*}

In order to go beyond the study of the spontaneous activity of the network, we consider its response to an external oscillatory signal.
While signal transmission in linear systems is fully characterized by the frequency response function of the system and by the noise spectrum of the output, the situation is different in the nonlinear neural network that we study here. 
Similarly to previous approaches \cite{Rajan10}, we provide oscillatory input to each unit in the microscopic network, randomizing the phase (Fig.~\ref{fig:driven_net}a)
\begin{equation}
I_i(t) = A_I\cos \left( 2\pi f_I t + \theta_i \right) \quad ,
\end{equation}
where $\theta_i \sim \mathrm{U}(0,2\pi)$.
The corresponding power spectral density of the input is given by $S_I(f) = \left(A_I^2/4\right) \cdot \left(\delta(f-f_I) + \delta(f+f_I) \right)$.
Thanks to the phase randomization, the network still reaches a stationary state and the mean $\langle x(t) \rangle$ remains at zero.
Notice that even if in this case the input is non-Gaussian, the mean-field equation for the power spectrum (Eq.~\ref{eq:spectrum_self_consistent-A}) is still valid.
However, since $x$ is also not Gaussian anymore, in order to find the mean-field solution we need to modify our iterative scheme by splitting the activation variable $x$ into its Gaussian and its oscillatory part \cite{Rajan10}.

The presence of the input affects the dynamics of the mean-field network, quantified by the power spectral density (Fig. \ref{fig:driven_net}b).
If the input is given while the network is in the chaotic regime ($g>g_c$), sharp peaks at the driving frequency $f_I$ and multiples thereof are elicited by the external input, standing out from a background power spectrum that is deformed compared to the case without the external input. 
For $f_I>f_p$, as in the example, the bumps of the background spectrum are slightly shifted toward larger values.
The opposite happens if $f_I<f_p$.
Notice that both this shift and the shaping of the chaotic activity are nonlinear effects due to the recurrent dynamics.
As an additional nonlinear effect, the network activity also exhibits harmonics at the driving frequency of the external input.

To characterize the response to the external stimulus, we split the power spectrum $S_x(f)$ into an oscillatory component and a chaotic component that constitutes the background activity
\begin{equation}\label{eq:osc_chaos_separation}
S_x(f) = S_{\text{bkg}}(f) + S_\text{osc}(f) \defeq S_\text{bkg} + \sum_{k=1}^\infty b_k \left(\delta(f-kf_I) + \delta(f+kf_I)\right) \quad ,
\end{equation}
where $b_k$ are positive coefficients and we included the multiples of the driving frequency in order to account for the harmonics. 
To solve the mean-field equations numerically, we have to consider a finite frequency bin $\Delta f$ (in our numerical results, $\Delta f=0.001$).
As a consequence, the heights of the delta peaks in the power spectrum in Eq.~\ref{eq:osc_chaos_separation} are finite and depend on $\Delta f$.
First, we will look at the transmission of the oscillatory signal near the driving frequency, i.e. how much of the peak in the power spectrum $S_x(f)$ at $f=f_I$ is due to the oscillatory drive and how much is due to the background activity.
At the driving frequency $f_I$  we write (see Fig. \ref{fig:driven_net}c)
\begin{equation}
S_x(f_I) = A_{\text{bkg}} + A_{\text{osc}} \defeq \frac{S_x(f_I-\Delta f) + S_x(f_I-\Delta f)}{2} + \frac{b_1}{\Delta f} \quad ,
\end{equation} 
i.e. we measure the contribution of chaotic activity to the power spectrum at the driving frequency by interpolating the power spectrum at neighboring frequencies.
The signal-to-noise ratio (SNR) at the driving frequency $f_I$ is then given by
\begin{equation}\label{eq:SNR-def}
\text{SNR}(f_I) = \frac{A_\text{osc}}{A_\text{bkg}} \quad .
\end{equation}
Notice the size of the frequency bin $\Delta f$ scales the SNR, but since we are interested in the dependency of the SNR on $f_I$ and not in its numerical value, this scaling factor can be neglected. 
Finally, we have seen in the example in Fig.~\ref{fig:driven_net}b that the oscillatory input can suppress background activity at frequencies far from $f_I$.
In order to quantify this chaos-suppression effect, we split the total variance of $x$ into two contributions (Fig.~\ref{fig:driven_net}c)
\begin{equation}
\text{Var}(x) = \int_{-\infty}^\infty S_{\text{bkg}}(f) df + 2 \sum_{k=1}^\infty b_k =: P_\text{bkg} + P_{\text{osc}} \quad .
\end{equation}

\begin{figure*}[t!]
\begin{center}
\includegraphics[scale=1]{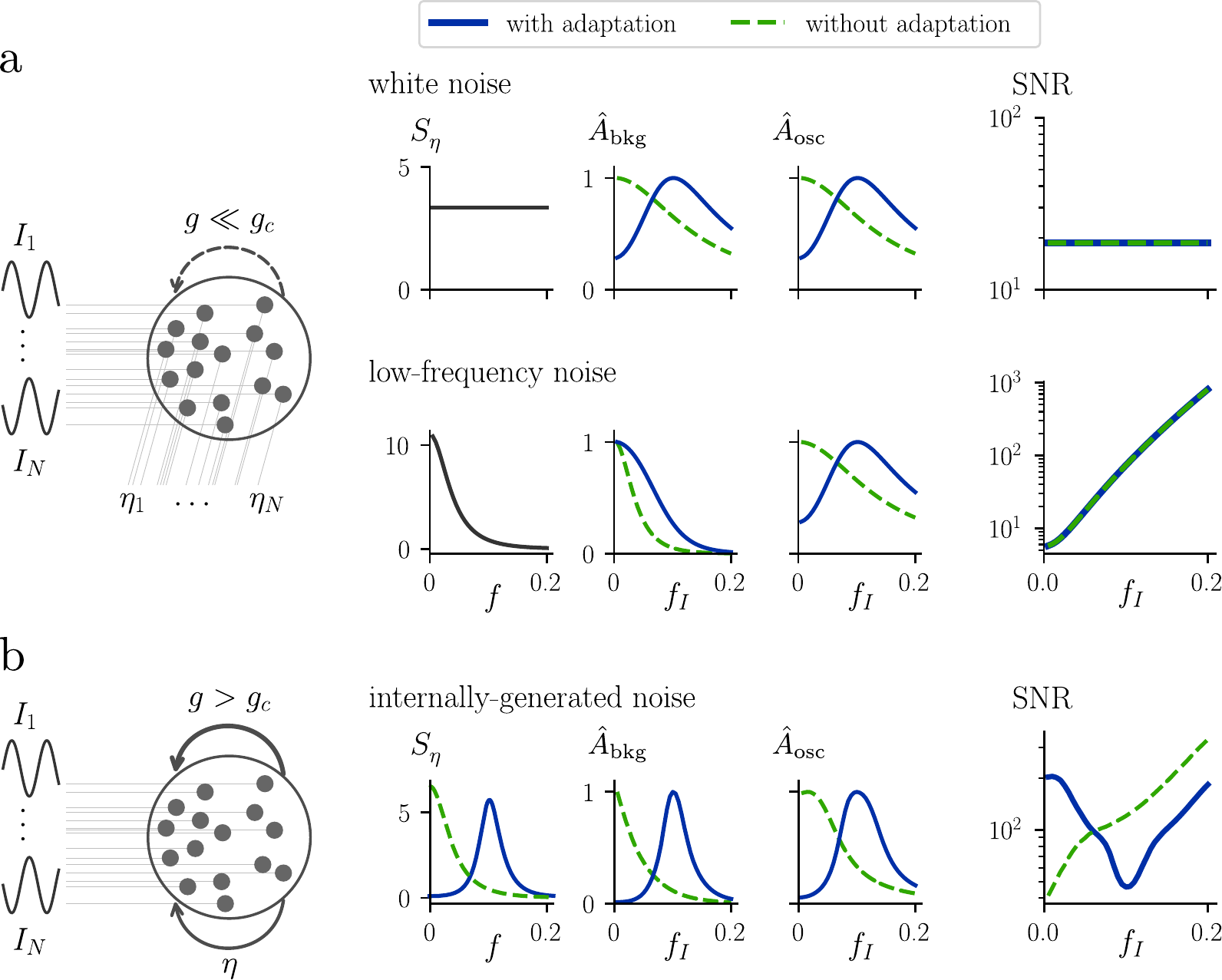}
\vspace{3mm}
\end{center}
\caption[Signal-to-noise ratio]{\textbf{Adaptation shapes the SNR in the chaotic regime.}
\textbf{a:} For small $g$, a recurrent network driven by an oscillatory input and external noise can be analyzed in the linear response theory framework.
Top row: response of the network to oscillatory drive and independent \textit{white} noise to each neuron.
Bottom row: response of the network to oscillatory drive and independent \textit{low-frequency} noise to each neuron.
For each row, from left to right, we plot the power spectrum of the input noise, the background component of the power spectrum $\hat{A}_\text{osc}$, the oscillatory component of the  power spectrum $\hat{A}_\text{osc}$, and the SNR as a function of the driving frequency.
The hat over the symbols $A_\text{bkg}$ and $A_\text{osc}$ indicates that, to highlight the network shaping, they are normalized to have the same maximum height (equal to one).
Notice that, since both signal and noise are shaped in the same way in the linear response framework, the introduction of adaptation does not affect the SNR.
\textit{Caption continues on the next page.}
\label{fig:SNR}
}
\end{figure*}
\addtocounter{figure}{-1}
\begin{figure*}[ht!]
\caption{\textit{Continues from previous page.}
\textbf{b:}
For large $g$, the network is subject to internally generated noise and driven by oscillatory input.
We plot the same quantities as in panel \textbf{a}.
Notice that, due to the nonlinearity of the network, signal and internally-generated noise are shaped in different ways, with the signal being subject to a broader effective filter.
As a consequence, the introduction of adaptation in the nonlinear network shapes the SNR by favoring low frequencies.
Parameters of the network with adaptation for both panels: $\gamma=0.25$, $\beta=1$ $g=2g_c(\gamma,\beta)$ and $A_I=0.5$.
}
\noindent\makebox[\linewidth]{\rule{\textwidth}{0.4pt}}
\end{figure*}

\subsubsection{Weak drive and signal transmission.}
If the oscillatory input is weak, chaos is not entirely suppressed and acts as internally-generated noise on the transmission of the oscillatory input.
We now study how the network transmits this oscillatory input signal, and how the transmission quality depends on the signal frequency $f_I$.
It is known from linear response theory that the transmission of weak signals through single homogeneous populations with strong (intrinsic or external) noise does not benefit from adaptation \cite{Schwalger13a,DegSch14}. This is because in the signal-to-noise ratio (SNR) both the signal and the noise are affected in the same way \cite{DegSch14}.  
We wondered whether in a strongly coupled, large random network, adaptation could have a different effect on the oscillatory signal than on the noise, thereby re-shaping the SNR.
A particularly interesting question is how signals are transmitted in the presence of purely intrinsically-generated chaotic fluctuations that are shaped by adaptation and recurrent connectivity.

To understand why adaptation cannot shape the SNR in a weakly-coupled network, consider our random network in the non-chaotic regime, with $g\ll g_c$ (Fig. \ref{fig:SNR}a).
If we drive the non-chaotic network with oscillatory input together with a noise source $\eta$, the typical response of one unit in the network can be approximated using the mean-field linear frequency response function $\tilde{\chi}_\beta(f)$, whose square modulus is given by (see appendix \ref{app:DMFT_FP_stability})
\begin{equation}
|\tilde{\chi}_\beta(f)|^2 = \frac{\tilde{G}(f)}{1-g^2 \tilde{G}(f)} \quad ,
\end{equation}
where we add the subscript $\beta$ to stress that $\tilde{\chi}_\beta$ depends on the adaptation parameters $\gamma$ and $\beta$ (cf. Eq.~\ref{eq:micro-dynamics-a}).
If we indicate by $S_I(f)$ and $S_\eta(f)$ the power spectral density of the oscillatory input and of the external noise respectively, the power spectral density of the output (taken as the network activity $x$) can be approximated by
\begin{equation}
S_x(f) = |\tilde{\chi}_\beta(f)|^2\left(S_I(f) + S_\eta(f) \right) \quad .
\end{equation}  
This means that both the signal and the noise are shaped by the same factor $|\tilde{\chi}_\beta(f)|^2$ that characterizes the network (Fig. \ref{fig:SNR}a).
The SNR of the output at the driving frequency, defined as in Eq.~\ref{eq:SNR-def}, is given by
\begin{equation}\label{eq:SNR-linear}
\mathrm{SNR}(f_I) = \frac{A_\text{osc}}{A_\text{bkg}} =
\frac{|\tilde{\chi}_\beta(f)|^2S_I(f_I)}{|\tilde{\chi}_\beta(f)|^2S_\eta(f_I)} = 
 \frac{S_I(f_I)}{S_\eta(f_I)} \quad ,
\end{equation}
i.e. the parameters of the network, reflected in the linear response function $\chi_\beta$, do not influence the SNR (Fig.~\ref{fig:SNR}a).
Notice that we considered the activation variable $x$ as our output.
We verified that considering instead the firing rate $\phi(x)$ as the output yields qualitatively the same results, therefore we will for simplicity continue our analysis for the output $x$.
Eq.~\ref{eq:SNR-linear} implies that the SNR depends only on the power spectra of the signal and of the noise.
For example, if we consider low-frequency dominated noise, high-frequency signals will be transmitted more easily, but once again the introduction of adaptation will not play any role (Fig.~\ref{fig:SNR}a).
While this argument is based on a linear response approximation, we verified using the DMFT solution that the linear approximation is quite accurate.
Deviations are visible very close to the criticality, but once again the SNR is almost entirely independent of the neuron parameters.

The findings are completely different for a network in the chaotic phase, i.e. $g>g_c$.
As discussed above, in this regime the network produces internal fluctuations whose power spectrum depends on single neuron parameters (Fig.~\ref{fig:SNR}b, see also section \ref{sec:chaotic-regime}).
For clarity, let us assume that there is no external noise, such that the noise is only internally generated by the network.
In this case, the linear response theory framework cannot be applied; in order to predict the effect of the network in shaping both the input and the internally-generated noise, we need to solve the DMFT equations (Eq.~\ref{eq:spectrum_self_consistent-A}) iteratively.
As in the previous section, the resulting power spectrum can be split into a chaotic component and into an oscillatory component (see Eq.~\ref{eq:osc_chaos_separation}).
How does the introduction of adaptation shape these two components?
We have seen that in the presence of adaptation the network can enter a state of resonant chaos that differs from the traditional chaos of a network without adaptation because of the presence of a dominant frequency band centered at $f_0$ (Fig.~\ref{fig:SNR}b, see also section \ref{sec:chaotic-regime}).
The state of resonant chaos survives in the presence of weak input.
As the driving frequency changes, the amplitude of the transmitted signal $A_\text{osc}$ passes through a maximum at the resonance frequency $f_0$ of the network; however, $A_\text{osc}$ decreases with the distance from the resonance frequency more slowly than the noise amplitude $A_\text{bgk}$ does (Fig.~\ref{fig:SNR}b).
This difference is related to the nonlinear sharpening effect of the recurrent network, that can be captured by DMFT.
As a consequence, the SNR is maximal at very slow driving frequencies, goes through a minimum at the resonance frequency $f_0$ before it increases again (Fig.~\ref{fig:SNR}b).
We can conclude that in the chaotic regime adaptation improves the SNR at low frequencies, whereas in weakly-coupled, non-chaotic networks such an improvement cannot be observed, independently of the choice of the adaptation parameters $\gamma,\beta$.
If the strength of the input is increased, the interaction between noise and signal becomes stronger, leading to a deformation of the SNR (Fig.~\ref{fig:varAmp}a).
However, even for strong drive we observe a peak of the SNR at frequencies that are lower than the resonance one.

\subsubsection{Strong drive and chaos suppression.}
\begin{figure*}[ht!]
\begin{center}
\includegraphics[scale=1]{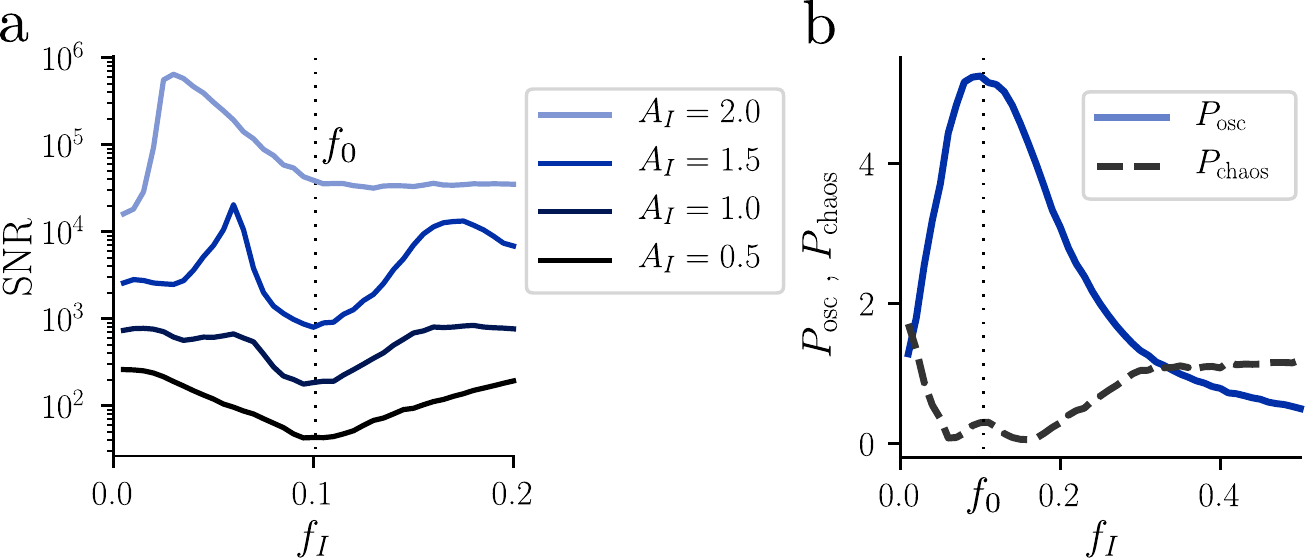}
\end{center}
\caption[Strong drive]{\textbf{Effect of a strong oscillatory input.}
\textbf{a:} SNR at the driving frequency $f_I$ as a function of the driving frequency, for different values of the signal amplitude $A_I$.
As $A_I$ increases, nonlinear interaction between signal and noise become stronger, leading to a qualitative change in the SNR profile.
\textbf{b:} Total power of the chaotic (black dashed) and oscillatory (light blue) components of the power spectrum, in the case of strong input ($A_I=1.5$).
For both panels, $\gamma=0.25$, $\beta=1.0$, and $g=2g_c(\gamma,\beta)$.
\label{fig:varAmp}
}
\end{figure*}

In the presence of strong input, chaos suppression together with the formation of a sharp peak are indications that at the microscopic level the network is driven towards a limit cycle.
Similarly to \cite{Rajan10}, we now study how chaos suppression depends on the driving frequency $f_I$.
By solving the DMFT equations (Eq.~\ref{eq:spectrum_self_consistent-A}) in the presence of external input, we find that both $P_\text{chaos}$ and $P_\text{osc}$ exhibit a non-monotonic dependence on $f_I$ (Fig. \ref{fig:varAmp}b).
$A_\text{osc}$ depends smoothly on $f_I$, reaching its largest value around $f_0$.
On the other hand, $A_\text{chaos}$ is zero for input frequencies that are close to $f_0$, indicating that the network is driven into a limit cycle.
While a network without adaptation also exhibits such a non-monotonic dependence \cite{Rajan10}, in our case this effect is more pronounced due to the resonant power spectrum of the spontaneous activity in the presence of adaptation.

\section{Discussion}\label{sec:discussion}

We studied how the dynamics of a random network of rate neurons are shaped by the properties of single neurons, and in particular by the presence of history-dependent mechanisms such as adaptation.
To this end, we generalized DMFT, a well-established theoretical tool \cite{Sompolinsky88}, to the case of multi-dimensional rate units.
This allowed us to reduce the high-dimensional, deterministic network model to a low dimensional system of stochastic differential equations.
Standard approaches to solving the mean-field equations \cite{Sompolinsky88} were not fruitful in the multi-dimensional setting.
However, the mean-field solution could be found efficiently in a semi-analytical way using an iterative approach.
The iterative approach highlights how recurrent connections sharpen the response function of single neurons, i.e. how bands of preferred frequencies become narrower (see also appendix \ref{app:qualitative}).
Previous studies that considered the role of single neuron properties on random network dynamics focused only on the role of the gain function \cite{Kadmon15,Mastrogiuseppe17}.
To our knowledge, this is the first result that relates the single neuron frequency response function  to the spectral properties of random network dynamics.

We studied in detail the important case of neuronal adaptation, using a two-dimensional rate model.
We showed that adaptation extends the stability region of a recurrent network of rate units because the transition from a stable fixed point to a fluctuating regime happens at $g=g_c > 1$, i.e. for higher coupling strength than for the network without adaptation.
Crucially, above the criticality and for slow adaptation, the dynamics settle in a state of ``resonant chaos'' that, unlike the chaotic activity of networks of rate units without adaptation, is dominated by a nonzero resonance frequency.
We observed that the resonance frequency can be computed from the single unit properties and it is therefore independent of the connection strength $g$. 
On the other hand, the presence of recurrent connections increases the coherence of the oscillations and therefore influences the correlation time.
The oscillation coherence is maximal at the onset of chaos and decreases with $g$, for $g>g_c(\gamma, \beta)$.
Indeed, as it is typical of critical behavior, the correlation time in the chaotic phase diverges when approaching the criticality.
In the presence of adaptation, this happens because the system approaches a limit cycle. 

It is interesting to observe that for slow adaptation there are two separate contributions to the correlation time of the network activity: an oscillatory component, related to the resonance frequency, and a long tail that scales with the adaptation timescale.
For finite $\tau_a$, the correlation time diverges when $g\rightarrow g_c$ due to the oscillatory component.
For $\tau_a\rightarrow\infty$, the correlation time also diverges, but this is due to the long tail, since both the resonance frequency and the Q-factor go to zero for large $\tau_a$, yielding a finite and therefore sub-dominant contribution to the correlation time.
Such multi-scale structure of the autocorrelation could be advantageous for network computations that require expressive dynamics over multiple timescales, as it is often the case in motor control.
Indeed, adaptation has been proposed to play a role in sequential memory retrieval \cite{Deco05}, slow activity propagation \cite{Setareh18}, perceptual bistability \cite{ShpMor09} and decision making \cite{Theodoni11}.
Moreover, SFA has beneficial consequences both for reservoir computing approaches \cite{Nicola17} and for spiking neuron-based machine learning architectures \cite{Bellec18}.
Further work could explore the relation between long correlation time induced by adaptation and computational properties.

In the presence of strong oscillatory input, chaos can be suppressed \cite{SchArs06,Rajan10}.
In particular, in the presence of adaptation, chaos is more easily suppressed when the driving frequency is close to the resonance frequency.
In the presence of weak input, chaos is not fully suppressed.
Interestingly, we found that in the chaotic regime the presence of adaptation shapes the SNR in frequency space.
In particular, adaptation increases the SNR for low-frequency signals, a possibly important feature since behaviorally relevant stimuli can have information encoded in slow signal components \cite{Melamed04}.
Crucially, this effect is not present in the sub-critical regime ($g<g_c$), since signal and external noise are shaped together \cite{DegSch14}.
It is known that the properties of biological neurons, including adaptation parameters, can be dynamically adjusted using neuromodulators \cite{McCormick89,Stiefel09}.
In view of our results, this would allow to dynamically shape the SNR depending on the requirements imposed by the behavioral context.

While our theory is applicable to single units with $D$ interacting variables, the effect of a single adaptation variable ($D=2$) on the dynamics of random recurrent networks was also studied independently and simultaneously by another group \cite{Beiran18}, who reached results consistent with ours \cite{Muscinelli18}.
The authors of \cite{Beiran18} used a slightly different network architecture and did not focus on the relation between single neuron response and spectral properties, but rather on the correlation time of the network activity and on the effect of white noise input.
One major difference is the conclusion reached regarding correlation time: by using a different definition, in \cite{Beiran18} the authors conclude that the correlation time \textit{does not} scale with the adaptation timescale.
Based on our analysis, we infer that the definition of correlation time used in \cite{Beiran18} captures only the oscillatory contribution to the correlation time, and not its long tail.

Current mean-field theories for spiking neural networks \cite{Bru00} are self-consistent only with respect to mean activities (firing rates), whereas second-order statistics such as autocorrelation function or power spectral density of inputs and outputs are inconsistent \cite{WieBer15}.
While iterative numerical procedures are available \cite{LerUrs06,DumWie14,WieBer15}, a self-consistent analytical calculation of the autocorrelation (or power spectrum) via DMFT for networks of spiking neurons is known to be a hard theoretical problem.
In the present manuscript, the rate-based modeling framework allowed us to put forward explicit expressions for the map of autocorrelations.
For a general nonlinearity $\phi(x)$, this map takes the form of an infinite series (according to Eq.~\ref{eq:infinite_series_x1} in appendix \ref{app:qualitative}).
However, for polynomial nonlinearities the series simplifies to a finite sum, e.g. Eq.~\ref{eq:cubic-map} in appendix \ref{app:nonlinearities}, which permits a closed-form analytical expression for the iterative map. Therefore, our study offers a unique method for the calculation of the autocorrelation in biologically constrained random neural networks, and thus represents a promising step towards a self-consistent mean-field theory beyond first-order rate models \cite{Sompolinsky88,Kadmon15}.

\subsection*{Extensions and generalizations}
We see four extensions to the work presented in the present manuscript.
First, our study is limited to rate neurons while it would be interesting to extend the analysis to spiking neuron models.
As a first step in this direction, previous work has already investigated the introduction of white noise in random rate networks \cite{Kadmon15,Schuecker18,Beiran18}, which would be straightforward to include in the case of $D$-dimensional rate units.
Second, our framework can readily be extended to multiple adaptation variables (see Fig.~\ref{fig:multi-d} for two examples).
This is a key feature in order to account for realistic SFA, which is known to have multiple timescales and it has been shown to have power-law structure \cite{Fairhall01,La-Camera06,Lundstrom08,Pozzorini13}.
Interestingly, our framework can be extended to power-law adaptation, since we require only the knowledge of the linear frequency-response function of the single neurons.
We expect that in this situation the internal noise generated by the network will also have a power-law profile of the type $f^\alpha$, with $\alpha>0$.
With such a noise spectrum, the signal that maximizes information transmission should be dominated by low-frequencies in a power-law fashion \cite{Rieke96,Pozzorini13}.
Third, the introduction of additional structure in the connectivity, such as low-rank perturbations \cite{Mastrogiuseppe18}, attractor structure \cite{Pereira18}, or large scale connectivity of the brain \cite{Deco11}, 
could give rise to interesting dynamics when combined with single units with multiple adaptation variables.
Finally, while our study focused on neural networks, random network models are used in other areas of biology and physics \cite{Mendes02}.
By extending mean-field theory techniques to more complex node dynamics, our approach also contributes to understanding the interaction between node dynamics and network structure in more general settings.

\section*{Acknowledgments}
The authors would like to thank Johanni Brea for helpful discussions and comments.
This research was supported by the Swiss national science foundation, grant agreement 200020\_165538.

\appendix

\section{Stability of the fixed point}\label{app:stability-microscopic}

\paragraph*{General theory.}
The system of $N\cdot D$ coupled nonlinear differential equations (Eq.~\ref{eq:multi-d-micro}) becomes intractable for large $N$.
However, because  $\phi(0)=0$, the system has a fixed point at the origin $\{x_i^\alpha=0\}_{i=1,\dotsc,N}^{\alpha=1,\dotsc,D}$, the stability of which can be studied owing to the clustered structure of the system.
The Jacobian at the fixed point is given by
\begin{equation}
\mathrm{B} =
\left( 
\begin{array}{c c c c}
\mathrm{A}^{11}\mathrm{I}_N + \phi'(0)\mathrm{J} & \mathrm{A}^{12}\mathrm{I}_N & \dots & \mathrm{A}^{1D}\mathrm{I}_N \\
\mathrm{A}^{21}\mathrm{I}_N  & \mathrm{A}^{22}\mathrm{I}_N & \dots & \mathrm{A}^{2D}\mathrm{I}_N \\
\dots & \dots & \dots & \dots\\
\mathrm{A}^{D1}\mathrm{I}_N  & \mathrm{A}^{D2}\mathrm{I}_N & \dots & \mathrm{A}^{DD}\mathrm{I}_N
\end{array}
\right) \quad ,
\end{equation}
where $\mathrm{J}$ is the random connectivity matrix and $\mathrm{I}_N$ is the $N$-dimensional identity matrix.
The matrix $\mathrm{B}$ is of size $ND \times ND$ and it therefore has $ND$ eigenvalues.
Since all the blocks of $\mathrm{B}$ commute with each other, we can apply the result of \cite{Silvester00} to find a relation between the eigenvalues of $\mathrm{J}$, $\mathrm{A}$ and $\mathrm{B}$
\begin{equation}\label{eq:multi-d-eigvals_transform}
\lambda_\mathrm{J} = \frac{\prod_{i=1}^D\left(\lambda_\mathrm{B}-\lambda_\mathrm{A}^i\right)}{\phi'(0)\prod_{j=1}^{D-1}\left(\lambda_{\mathrm{B}}-\lambda_{\mathrm{A}^-}^j\right)} \quad ,
\end{equation}
where $\mathrm{A}^-$ is the matrix obtained by removing the first column and the first row from the matrix $\mathrm{A}$.
This expression is valid for all the eigenvalues of $\mathrm{B}$ that are not coincident with those of $\mathrm{A}^-$.
Eq.~\ref{eq:multi-d-eigvals_transform} can be transformed into a polynomial equation of degree $D$ in $\lambda_\mathrm{B}$, so that for every value of $\lambda_\mathrm{J}$ we obtain $D$ eigenvalues of $\mathrm{B}$, as expected.
From now on we will assume that, without loss of generality, $\phi'(0)=1$.

In the $N\rightarrow \infty$ limit, the eigenvalues $\lambda_\mathrm{J}$ are known to be uniformly distributed on a disk in the complex plane, centered at zero and of radius $g$ \cite{Girko85}.
If one can invert Eq.~\ref{eq:multi-d-eigvals_transform}, it becomes computationally fast to compute the eigenvalues of the Jacobian in the $N\rightarrow\infty$ limit without finite-size effects.
Whether one can obtain an explicit inverse formula depends on the dimensionality and on the entries of the matrix $\mathrm{A}$.

\paragraph*{Network with adaptation.} 

\begin{figure*}[!ht]
\begin{center}
\includegraphics[width=1\textwidth]{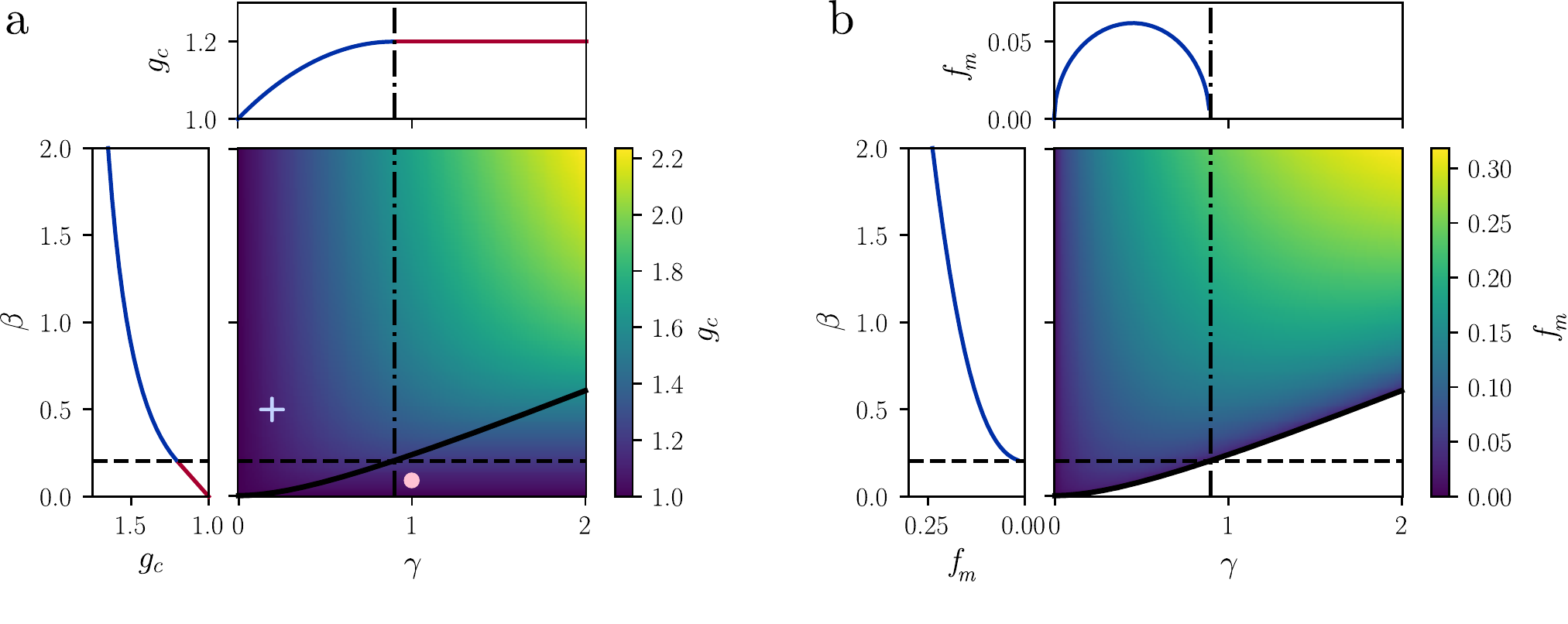}
\end{center}
\caption[Stability of the fixed point and local properties]{\textbf{Stability of the fixed point and local properties.}
\textbf{a:} Critical value of the coupling $g_c$ (color code, right) for different adaptation parameters $\gamma$ (horizontal axis) and $\beta$ (vertical axis).
The curve $\beta_H(\gamma)$ (solid black line) separates the regions of the $\gamma - \beta$ plane in which for increasing $g$ we encounter a Hopf bifurcation (above $\beta_H(\gamma$)) or a saddle-node bifurcation (below $\beta_H(\gamma)$).
Cross and filled circle: parameters used in Fig. \ref{fig:microscopic}.
Left inset: dependence of $g_c$ on $\beta$ for fixed $\gamma=0.9$.
Top inset: dependence of $g_c$ on $\gamma$ for fixed $\beta=\beta_H(\gamma=0.9)$.
Blue line: Hopf bifurcation; red line: saddle-node bifurcation.
\textbf{b:} Resonance frequency $f_0$ for different adaptation parameters $\gamma, \beta$.
Notice that in the non-resonant region the resonance frequency is not defined.
Left inset: square-root increase of $f_m$ as a function of $\beta$ for fixed $\gamma=0.9$.
Top inset: non-monotonic behavior of $f_m$ as a function of $\gamma$, for fixed $\beta=\beta_H(\gamma=0.9)$.
}
\label{fig:stability}
\end{figure*}

For the two-dimensional model defined by Eqs.~(\ref{eq:micro-dynamics-x}, \ref{eq:micro-dynamics-a}), we can invert Eq.~\ref{eq:multi-d-eigvals_transform}, and obtain an expression for the eigenvalues of the Jacobian  
\begin{equation}\label{eq:eigenvalue_transform}
\lambda_{\mathrm{B}}(\lambda_\mathrm{J}) = \frac{1}{2}\left( -1 -\gamma +\lambda_{\mathrm{J}} \pm \sqrt{(\lambda_{\mathrm{J}} -1 +\gamma)^2 - 4\gamma\beta} \right).
\end{equation}
Using the mapping between the eigenvalues of the connectivity matrix $\mathrm{J}$ and those of the Jacobian matrix $\mathrm{B}$ ( Eq.~\ref{eq:eigenvalue_transform}), we find the critical value of $g$ for which the stability of the fixed point is lost
\begin{equation}\label{eq:g_critical-app}
g_c(\gamma,\beta) = 
\begin{cases}
\sqrt{1-\gamma(\gamma+2\beta) + 2\sqrt{\gamma^2\beta(2\gamma+2\beta+2)}},&\beta > \beta_H(\gamma) \\
1+\beta,&\beta\leq \beta_H(\gamma)
\end{cases}
\end{equation}
where $\beta_H(\gamma) = -1 -\gamma + \sqrt{2\gamma^2 + 2\gamma +1}$.
The critical value $g_c$ can also be calculated from dynamical mean-field theory (see appendix \ref{app:DMFT_FP_stability}).

The bifurcation that characterizes the loss of stability depends on two parameters, viz. the ratio of timescales $\gamma$ and the strength of the adaptation $\beta$.
To further characterize the bifurcation at $g=g_c(\gamma,\beta)$, we can study the imaginary part of the critical eigenvalue, i.e. the one with real part equal to zero at $g=g_c(\gamma,\beta)$.
If the adaptation strength $\beta$ has a value $\beta\leq \beta_H(\gamma)$, then the imaginary part of the critical eigenvalue is equal to zero corresponding to a saddle-node bifurcation at $g=g_c(\gamma,\beta)$.
On the other hand, if $\beta>\beta_H(\gamma)$, then the critical eigenvalue is a pair of complex-conjugate, purely imaginary eigenvalues, a signature of a Hopf bifurcation.
Therefore, we introduce the curve $\beta=\beta_H(\gamma)$, which separates the positive quadrant of the $\gamma-\beta$ plane in two regions: one in which the system becomes unstable at the critical value $g_c(\gamma,\beta)$ via a saddle-node bifurcation, and another one in which the instability occurs via a Hopf bifurcation (Fig \ref{fig:stability}a). 
In the Hopf-bifurcation region, the imaginary part of the critical eigenvalues can be computed analytically:
\begin{equation}\label{eq:resFreq_micro}
\mathrm{Im}(\lambda_\mathrm{B}^c) = \sqrt{-\gamma^2 + \sqrt{\beta\gamma^2(\beta+2\gamma+2)}} =: 2\pi f_m \quad .
\end{equation}
The parameter $f_m$ is the frequency of low-amplitude oscillations close to the bifurcation, if $N<\infty$.
In the finite-$N$ case, we find numerically that these low-amplitude oscillations are stable.
When $N\rightarrow \infty$, however, we find that chaotic dynamics onset right above the bifurcation (see section \ref{sec:chaotic-regime}).
The frequency $f_m$ is monotonic in $\beta$ but non-monotonic in $\gamma$ (Fig. \ref{fig:stability}b), indicating that a slower adaptation variable (smaller $\gamma$) does not necessarily correspond to slower oscillations.
When considering codimension-two bifurcations, we have that for $g=g_c$ and $\beta=\beta_H(\gamma)$ the system undergoes a Bogdanov-Takens bifurcation.

\section{Qualitative study of the iterative map}\label{app:qualitative}

\begin{figure*}[!ht]
\begin{center}
\includegraphics[width=1\textwidth]{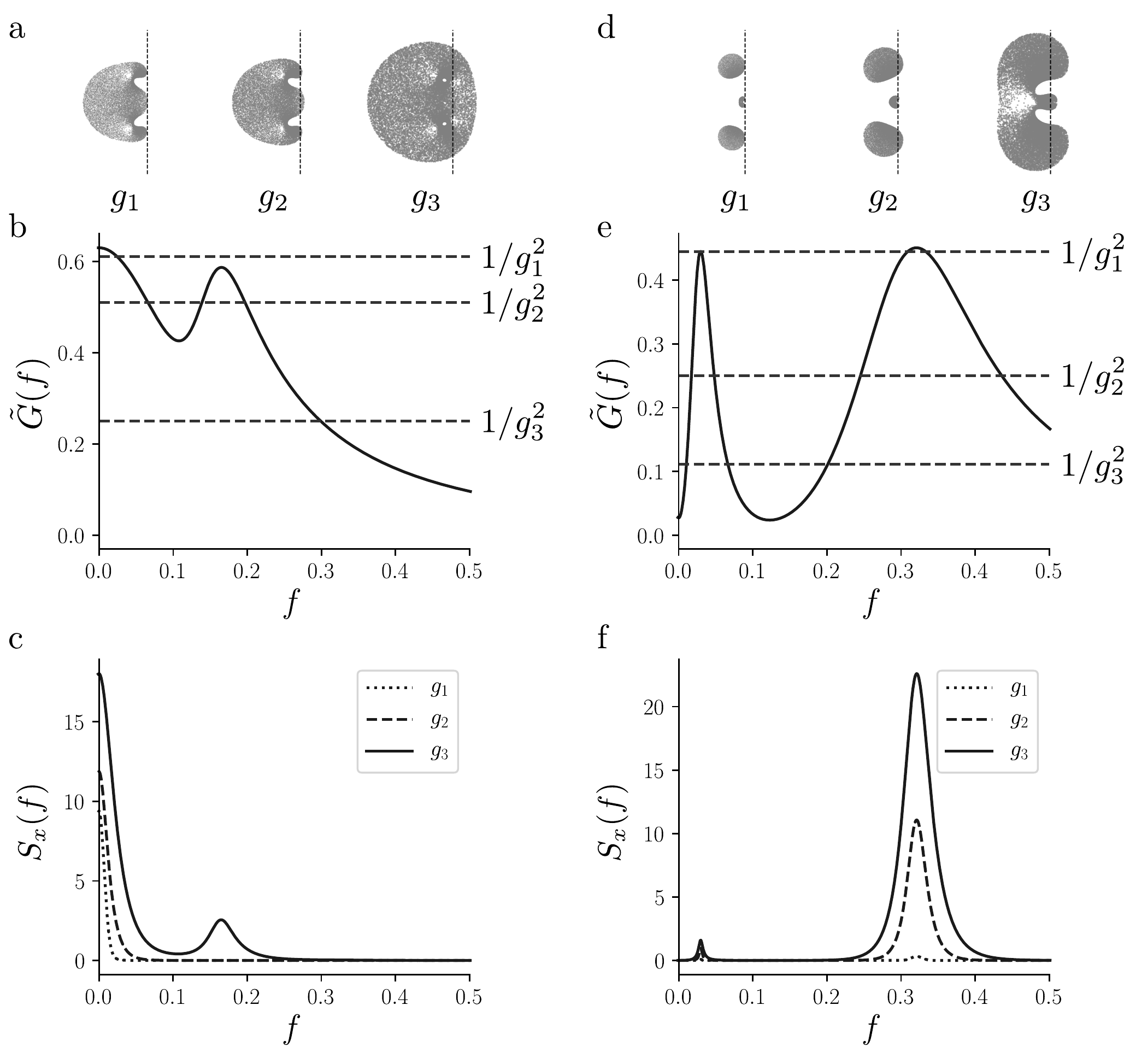}
\end{center}
\caption[Two examples of multi-dimensional rate models]{\textbf{Two examples of multi-dimensional rate models}
\textbf{a-b-c:} Analysis of a three-dimensional rate model.
Eigenvalue spectra (\textbf{a}) corresponding to the coupling values  $g_1= 1.28$, $g_2=1.4$ and $g_3 = 2$.
The dashed line indicates the imaginary axis.
In \textbf{b} we plot the linear response function of the single unit $\tilde{G}(f)$ (solid line), and the instability threshold corresponding to the three coupling values $g_1$, $g_2$ and $g_3$ (dashed lines).
In \textbf{c} we plot the solution of the mean field theory obtained with the iterative method for the three values of $g$, $g_1= 1.5$, $g_2=2$ and $g_3 = 3$.
\textbf{d-e-f:} Same as \textbf{a},\textbf{b},\textbf{c}, but for a four-dimensional rate model.
}
\label{fig:multi-d}
\end{figure*}

\begin{table}
\begin{center}
\begin{tabular}{c c}
Fig. \ref{fig:multi-d}a,b,c & Fig. \ref{fig:multi-d}d,e,f  \\
 \toprule \\

 $ \left(
 \begin{array}{c c c}
 -1 & -1 & -1 \\
 0.1 & -0.1 & 1.7 \\
 0.1 & -0.4 & -0.5
 \end{array} 
 \right)
 $
 &
$ \left(
 \begin{array}{c c c c}
 -1 & -1 & -1 & -1 \\
 1 & -0.5 & -0.65 & -0.6\\
 1 & 0.35 & -0.05 & -0.57 \\
 1 & 0.35 & 0.28 & -0.005
 \end{array} 
 \right)
 $
 \\
 \vspace{0.05cm} \\
\hline
\end{tabular}
\end{center}
\caption[Parameters of the models in the examples]{\textbf{Parameters of the models in Fig. \ref{fig:multi-d}.}
Matrix $\mathrm{A}$ defining the rate model for the different examples in Fig. \ref{fig:multi-d}.
}
\label{table:examples}
\end{table}

For a qualitative understanding of the effect of the iterations on the power spectral density, we exploit the fact that $x^1$ is a Gaussian process, for which the following formula holds \cite{Mal78}
\begin{equation}\label{eq:infinite_series_x1}
C_{\phi(x^1)}(\tau) = \sum_{n=0}^\infty \frac{1}{n!}\left(\left\langle \frac{d^n \phi}{d(x^1)^n}\right\rangle\right)^2 C_{x^1}^n(\tau) \quad ,
\end{equation}
where the angular brackets indicate the mean over the statistics of $x^1$.
Eq.~\ref{eq:infinite_series_x1} gives the effect of a nonlinearity $\phi$ on a the autocorrelation of a Gaussian process $x^1$.
By truncating the series after the first term, we get
\begin{equation}
C_{\phi(x^1)}(\tau) \simeq \left(\left\langle \phi'(x^1)\right\rangle\right)^2 C_{x^1}(\tau) \quad .
\end{equation}
Fourier transforming this equation we get an approximation of the power spectral density of $\phi(x^1)$
\begin{equation}
S_{\phi(x^1)}(f) \simeq \Psi_1\left( \int_{-\infty}^\infty S_x(f')df'\right) S_x(f) \quad ,
\end{equation}
where we we introduced the function  $\Psi_1\left( \int_{-\infty}^\infty S_x(f')df'\right):=\left(\left\langle \phi'(x^1)\right\rangle\right)^2$ to highlight the fact that the coefficient that multiplies $S_x(f)$ depends on the area under the power spectral density, i.e. on the variance of $x^1$, and is therefore nonlocal in frequency space.
We stress that retaining only the first term in Eq.~\ref{eq:infinite_series_x1} is different than considering a linear approximation of $\phi$, since the dependence of the coefficient on the variance would not appear in that case.

Using this approximation, we can express the power spectral density at the $n^{\text{th}}$ iteration of the iterative method, as a function of the initial power spectral density $S_{\phi(x^1)}^{(0)}(f)$ from which we started to iterate.
We obtain
\begin{equation}
(S_x)^{(n)}(f) = \left(\prod_{k=1}^{n-1}\Psi_1^{(k)} \right) \left(g^2 \tilde{G}(f)\right)^n S_{\phi(x^1)}^{(0)}(f) \quad ,
\end{equation} 
where $ \Psi_1^{(n)} := \Psi_1\left( \int_{-\infty}^\infty (S_x)^{(n)}(f')df'\right)$.
If we take $S_{\phi(x^1)}^{(0)}(f)$ to be constant and we define $a_{n} = \left(\prod_{k=1}^{n-1}\Psi_1^{(k)} \right)$, we can rewrite the above expression as
\begin{equation}\label{eq:S_n-simplified}
(S_x)^{(n)}(f) = a_n \left(g^2 \tilde{G}(f)\right)^n \quad .
\end{equation}
If $g>g_c$, there will be a range of frequencies for which $g^2 \tilde{G}(f)>1$, which implies that its $n^{\text{th}}$ power diverges when $n$ grows.
In a purely linear network, this phenomenon would lead to a blow-up of the power spectral density, in agreement with the fact that activity in a linear network is unbounded for $g>g_c$.
If $\phi$ is a compressive nonlinearity however, the coefficient $a_n$ will tend to zero for growing $n$, counterbalancing the unbounded growth of $\left(g^2 \tilde{G}(f)\right)^n$.
Based on Eq.~\ref{eq:S_n-simplified}, we would predict that all the modes for which $\tilde{G}(f)>1/g^2$ will get amplified over multiple iterations, while all the other modes will get suppressed.
While this is a highly simplified description, the suppression and the amplification of modes is clearly visible when observing the evolution of the power spectrum over iterations (Fig.~\ref{fig:mean_field_regimes}f) and when comparing the dynamics of the self-consistent solution (Fig. \ref{fig:multi-d}c,f) to the corresponding linear response function (Fig. \ref{fig:multi-d}b,e).
When truncating the series after the first order however, the mean-field network does not admit a self-consistent solution, for which we need to retain also higher order terms.
Such terms will balance the progressive sharpening of the power spectrum, allowing for a self-consistent solution. 

As an example of higher-order term, consider the next term in the series in Eq.~\ref{eq:infinite_series_x1}, given by
\begin{equation}\label{eq:infinite_series_second_order}
\begin{aligned}
&\frac{1}{2}\left(\left\langle \phi''(x^1)\right\rangle\right)^2 \left(C_{x^1}(\tau)\right)^2 \xrightarrow{FT} \\
\xrightarrow{FT}
&\frac{1}{2}\Psi_2 \left( \int_{-\infty}^\infty S_x(f')df'\right) \left( S_x \ast S_x \right)(f)
\end{aligned}
\end{equation}
where $\Psi_2$ is defined analogously to $\Psi_1$.
In general, higher-order terms will contain convolutions of the power spectral density with itself, which are responsible for the creation of higher harmonics.
To qualitatively understand this effect, consider the case in which $S_x(f)$ is a Dirac $\delta$-function with support in $f_0$.
In this case, the two-fold convolution of $S_x(f)$ with itself is again equal to a Dirac $\delta$-function, but centered in $2f_0$. 
A similar argument can be given for resonant power spectral densities, which implies that a self-consistent solution should exhibit harmonics of the fundamental resonance frequency.
Note that in this paper we considered odd functions, for which only odd terms in the series are nonzero.

For higher values of $g$, the relative importance of higher-order terms in the series in Eq.~\ref{eq:infinite_series_x1} will increase, leading to a broader power spectrum.
The self-consistent power spectrum however, seems to be always narrower than the single neuron linear response function.
For a possible explanation of this phenomenon, we consider the $g\rightarrow\infty$ limit, which was already studied in \cite{Crisanti18} for the network without adaptation.
Using the same technique, we conclude that in this limit the autocorrelation decay tends to be the same as one obtained for a single unit driven by white noise \cite{Crisanti18}.
In the frequency domain, this is equivalent to say that the power spectral density of the network tends to the one of a single unit driven by white noise.

\section{Mean-field theory derivation}\label{app:DMFT_derivation}

In this section, we extend the derivation of dynamical mean-field theory (DMFT) to the case of the network of multi-dimensional rate units.
Since there are no additional complication with respect to the standard case, we report here only the main steps.
For a review of the path-integral approach to DMFT, see e.g. \cite{Schucker16a,Crisanti18}.
The moment-generating functional corresponding to the microscopic system in Eq.~\ref{eq:multi-d-micro} is given by
\begin{align}
Z[\ve{j},\tilde{\ve{j}}](\mathrm{J}) = \int{\gaussd{\ve{x}}\gaussd{\tilde{\ve{x}}}
\exp\left[S_0[\ve{x},\tilde{\ve{x}}] - (\tilde{\ve{x}}^1)^T \mathrm{J} \phi(\ve{x^1(t)}) +\ve{j}^T\ve{x} +\tilde{\ve{j}}^T \tilde{\ve{x}} \right]  } \quad ,
\end{align}
where
\begin{equation}
S_0[\ve{x},\tilde{\ve{x}}] \defeq \tilde{\ve{x}}^T (\mathrm{I}_D\partial_t - \mathrm{A})\ve{x}
\end{equation}
and we introduced the notation $\tilde{\ve{x}}^T \ve{x} = \sum_\alpha \sum_i \int{\tilde{x}_i^\alpha(t)x_i^\alpha(t)dt}$.
The integral is over paths and bold symbols indicate vectors, over both the network space and the rate model space, so that $\gaussd{\ve{x}} \defeq \prod_\alpha \prod_i\gaussd{x_i^\alpha}$. 

We are interested in properties that are independent of the particular realization of the coupling matrix $\mathrm{J}$.
In order to extract those properties, we average over the quenched disorder by defining the averaged generating function
\begin{equation}
\bar{Z}[\ve{j},\tilde{\ve{j}}] \defeq \int \prod_{ij} dJ_{ij}\mathcal{N}\left(0, \frac{g^2}{N}; J_{ij} \right)
Z[\ve{j}^x,\tilde{\ve{j}}^x](\mathrm{J}) \quad .
\end{equation}
The average over each $J_{ij}$ can be computed by noticing that the terms corresponding to different $J_{ij}$ factorize and the integral can be solved by completing the square.
Since the details of this calculation are analogous to the one-dimensional case, we directly report the result 
\begin{align}
\bar{Z}[\ve{j}^x,\tilde{\ve{j}}^x] = \int{\gaussd{\ve{x}}\gaussd{\tilde{\ve{x}}}
\exp\left[S_0[\ve{x},\tilde{\ve{x}}] 
+\ve{j}^T\ve{x} +\tilde{\ve{j}}^T \tilde{\ve{x}} \right]} \,\times \nonumber \\
\times \exp\left[\frac{1}{2}\int_{-\infty}^{\infty}\left(\sum_i \tilde{x}_i^1(t)\tilde{x}_i^1(t') \right)
\left(\frac{g^2}{N}\sum_j \phi(x_j^1(t))\phi(x_j^1(t')) \right)dtdt'   \right] \quad .
\end{align}
We now aim to decouple the interaction term in the last line by introducing the auxiliary field
\begin{equation}
Q_1(t,s) \defeq \frac{g^2}{N} \sum_j \phi(x_j^1(t))\phi(x_j^1(s)) \quad .
\end{equation}
We introduce $Q_1$ in the generating functional by inserting the following representation of the unity
\begin{equation}
\int \gaussd{Q_1}\delta\left[-\frac{N}{g^2}Q_1(s,t) + \sum_j \phi(x_j^1(s))\phi(x_j^1(t)) \right] \quad ,
\end{equation}
where $\delta [\cdot]$ is the delta functional.
Using the integral representation of the delta functional leads to the introduction of a second auxiliary field, which we call $Q_2$. 
We obtain
\begin{align}
\bar{Z}[\ve{j}^x,\tilde{\ve{j}}^x] &= \int{\gaussd{Q_1}\gaussd{Q_2}\gaussd{\ve{x}}\gaussd{\tilde{\ve{x}}}
\exp\left[S_0[\ve{x},\tilde{\ve{x}}] 
+\ve{j}^T\ve{x} +\tilde{\ve{j}}^T \tilde{\ve{x}} \right]} \, \times \nonumber \\ 
&\exp\Biggl[\frac{1}{2}\int_{-\infty}^{\infty}\Biggl(\sum_i \tilde{x}_i^1(t)Q_1(t,t')\tilde{x}_i^1(t')+ 
\sum_i \phi(x_i^1(t) Q_2(t,t')\phi(x_i^1(t')) + \nonumber\\
&-\frac{N}{g^2}Q_1(t,t')Q_2(t,t') \Biggr)
dtdt'   \Biggr] \quad .
\end{align}
This expression has the advantage that any interaction between different units is removed and all the contribution coming from different units factorize.
It is convenient to rewrite the averaged generating functional as a field theory for two auxiliary fields $Q_1, Q_2$, i.e. we remove the vectorial response terms $\ve{j}^T\ve{x},\tilde{\ve{j}}^T \tilde{\ve{x}}$ and we add two scalar response terms for the auxiliary fields.
The result is
\begin{align}
\bar{Z}[j,\tilde{j}] = \int \gaussd{Q_1}\gaussd{Q_2} \exp\left(-\frac{N}{g^2}Q_1^T Q_2 + N \ln{Z[Q_1,Q_2]} + j^T Q_1 + \tilde{j}^T Q_2 \right) \nonumber \\
Z[Q_1,Q_2] \defeq \int{\gaussd{\ve{x}} \gaussd{\tilde{\ve{x}}} \exp\left( S_0[\ve{x},\tilde{\ve{x}}] +
\frac{1}{2} (\tilde{x}^1)^T Q_1 \tilde{x}^1 + \phi(x^1)^T Q_2 \phi(x^1) \right)} \quad , 
\end{align}
where we extended our notation to $ Q_1^T Q_2 \defeq \int\int Q_1(s,t)Q_2(s,t)dsdt$.
The crucial observation to make is that essentially all factors associated to different units factorized yielding the factor $N$.
For this reason, the integration is now only over all rate model indices but over only one unit index.
The remainder is the problem of one unit, characterized by $D$ variables, interacting with two external fields $Q_1,Q_2$.

The final step is to perform a saddle-point approximation, i.e. replace $Q_1, Q_2$ by their values that make the action stationary.
To do this, we need to solve the two saddle-point equations
\begin{equation}
\frac{\delta }{\delta Q_{\{1,2\}}}\left(\frac{N}{g^2}Q_1^T Q_2 +N\ln Z[Q_1,Q_2] \right) = 0
\end{equation}
These equations are analogous to the ones in the one-dimensional case, and lead to the saddle-point solution
\begin{equation}
\begin{aligned}
&Q_1^*(s,t) = g^2 C_{\phi(x^1)}(s,t) \\
&Q_2^*(s,t) = 0
\end{aligned}
\quad ,
\end{equation}
where $C_{\phi(x^1)}(s,t)$ is the autocorrelation function of $\phi(x^1)$ evaluated at the saddle point solution.
The averaged generating functional at the leading order in $N$ can be written as
\begin{equation}
\bar{Z}^* \propto \int \gaussd{\ve{x}}\gaussd{\tilde{\ve{x}}} \exp\left( S_0[\ve{x},\tilde{\ve{x}}]
+ \frac{g^2}{2} (\tilde{x}^1)^T C_{\phi(x^1)} \tilde{x}^1 \right) \quad .
\end{equation}
This is the statistical field theory corresponding to $D$ linearly interacting variables, with $x^1$ that receives a Gaussian noise whose autocorrelation is given by $C_{\phi(x^1)}$.
Writing the corresponding differential equations results in our mean-field description (Eq. \ref{eq:multi-d-mean-field}).

\section{Mean-field theory with heterogeneous adaptation}\label{app:heterogeneous}
In this section, we will extend the derivation of the dynamic mean-field theory (DMFT) for the case of the network with heterogeneous adaptation.
We consider the case in which each neuron has different parameters, sampled i.i.d from the same distributions, and different parameters of the same neuron are uncorrelated with each other.
More precisely, we sample the elements of the matrix $\mathrm{A}_i$ for neuron $i$ as
\begin{equation}
\mathrm{A}_i^{\alpha\beta} \sim \mathcal{N}\left(\bar{\mathrm{A}}^{\alpha\beta},(\sigma^{\alpha\beta})^2\right) \quad ,
\end{equation} 
where the subscript $i$ runs over the neurons in the network.

In deriving the mean-field theory, most of the steps are identical to those in appendix \ref{app:DMFT_derivation}, so we will focus on the additional terms due to the new source of disorder.
We separate the contribution of mean adaptation parameters $\bar{\mathrm{A}}^{\alpha\beta}$ from the deviations, so that the generating functional reads
\begin{equation}
\begin{aligned}
Z[\ve{j},\tilde{\ve{j}}](\mathrm{J}) = \int{\gaussd{\ve{x}}\gaussd{\tilde{\ve{x}}}
\exp\left[S_0[\ve{x},\tilde{\ve{x}}] - (\tilde{\ve{x}}^1)^T \mathrm{J} \phi(\ve{x^1(t)}) - \sum_k\ve{x}_k^T(\mathrm{A}_k-\bar{\mathrm{A}})\ve{x}_k +\ve{j}^T\ve{x} +\tilde{\ve{j}}^T \tilde{\ve{x}} \right]  } \, ,
\end{aligned}
\end{equation}
where
\begin{equation}
S_0[\ve{x},\tilde{\ve{x}}] \defeq \tilde{\ve{x}}^T (\mathrm{I}_D\partial_t - \bar{\mathrm{A}})\ve{x}
\end{equation}
and $\bar{\mathrm{A}}$ is the matrix of the expected values of $\mathrm{A}$.

The action $S_0$ is the same as for the network without heterogeneity, and when averaging over the connectivity disorder, we obtain the same result as for homogeneous network.
In this case however, we need to also average over the disorder due to heterogeneity, i.e. over all the $\mathrm{A}_k^{\alpha\beta}$.
The averaged generating functional will then result from the average
\begin{equation}
\bar{Z}[\ve{j},\tilde{\ve{j}}] := \int \left(\prod_{ij} dJ_{ij}\mathcal{N}\left(0, \frac{g^2}{N}; J_{ij} \right)\right)
\left(\prod_{\alpha\beta k} d\mathrm{A}_k^{\alpha\beta}\mathcal{N}\left(\bar{\mathrm{A}}_k^{\alpha\beta}, \sigma^{\alpha\beta}; \mathrm{A}^{\alpha\beta} \right)\right)
Z[\ve{j},\tilde{\ve{j}}](\mathrm{J}) \, .
\end{equation}
The new terms due to the heterogeneity result in integrations of the type
\begin{equation}
\frac{1}{\sqrt{2\pi (\sigma^{\alpha\beta})^2}}\int \exp \left(-\frac{1}{2(\sigma^{\alpha\beta})^2}\left(\mathrm{A}^{\alpha\beta}-\bar{\mathrm{A}}^{\alpha\beta}\right)^2 -\left(\mathrm{A}^{\alpha\beta}-\bar{\mathrm{A}}^{\alpha\beta}\right) \int \tilde{x_i}^{\alpha}(t)x_i^\beta(t)dt \right) \quad ,
\end{equation}
that can be solved by completing the square.
After averaging over both the connectivity disorder and the heterogeneity disorder, the generating functional reads 
\begin{equation}
\begin{aligned}
\bar{Z}[\ve{j}^x,\tilde{\ve{j}}^x] =& \int{\gaussd{\ve{x}}\gaussd{\tilde{\ve{x}}}
\exp\left[S_0[\ve{x},\tilde{\ve{x}}] 
+\ve{j}^T\ve{x} +\tilde{\ve{j}}^T \tilde{\ve{x}} \right]} \,\times \nonumber \\
&\times \exp\left[\frac{1}{2}\int\left(\sum_i \tilde{x}_i^1(t)\tilde{x}_i^1(t') \right)
\left(\frac{g^2}{N}\sum_j \phi(x_j^1(t))\phi(x_j^1(t')) \right)dtdt'   \right] \, \times \\
&\times \, \exp\left[\sum_{i\alpha\beta}\frac{(\sigma^{\alpha\beta})^2}{2}\int \tilde{x}_i^{\alpha}(t)x_i^\beta(t)x_i^\beta(t')\tilde{x}_i^\alpha(t')\right] \quad .
\end{aligned}
\end{equation}
The last term, which is due to the heterogeneity, factorizes into the contributions associated to different units.
From this point on, in order to derive the mean-field equations, we follow exactly the same steps as in appendix \ref{app:DMFT_derivation}, so we do not report those steps here.
The mean-field equations read
\begin{equation}
\dot{x}^\alpha(t) = \sum_{\beta=1}^D \left( \bar{\mathrm{A}}^{\alpha\beta} x^\beta(t)  +\eta_H^{\alpha\beta}(t)\right) 
                     + \delta^{\alpha 1}\left( \eta(t) +I(t)\right) \quad , 
\end{equation}
where $\eta_H^{\alpha\beta}$ are Gaussian processes associated to the heterogeneity, that all have mean zero and autocorrelation
\begin{equation}
\langle \eta_H^{\alpha\beta}(t)\eta_H^{\alpha\beta}(s)\rangle = (\sigma^{\alpha,\beta})^2 \langle x^\beta(t)x^\beta(s) \rangle \quad .
\end{equation}

For the particular case of adaptation with heterogeneity on the parameter $\beta$, as studied in section \ref{sec:chaotic-regime}, we have the following mean-field equations
\begin{align}
\dot{x}(t) =& -x(t) -a(t) + \eta(t) +I(t) \label{eq:mean-field-hetero-x}\\
\dot{a}(t) =& -\gamma a(t) + \gamma\bar{\beta} x(t) + \gamma\eta_H(t)\quad , \label{eq:mean-field-hetero-a} 
\end{align}
where $\eta_H(t)$ is a Gaussian process with mean zero and autocorrelation
\begin{equation}
\langle \eta_H(t)\eta_H(s)\rangle = \sigma_\beta^2 \langle x(t)x(s) \rangle \quad .
\end{equation}
From Eqs.~\ref{eq:mean-field-hetero-x},\ref{eq:mean-field-hetero-a}, we can find the self-consistent equation for the power spectrum:
\begin{equation}
S_x(f) = \tilde{G}_H(f) \left(g^2 S_{\phi(x)}(f) + S_I(f)\right) \quad ,
\end{equation}
where $\tilde{G}_H(f)$ is an effective filter given by
\begin{equation}
\tilde{G}_H(f) = \frac{\tilde{G}(f)}{1-\frac{\gamma^2\sigma_\beta^2}{\gamma^2 + \omega^2}\tilde{G}(f)} \quad ,
\end{equation}
where $\omega=2\pi f$.
The effective filter $\tilde{G}_H(f)$ predicts a larger power at low frequencies, similar to what is observed in simulations (cf. Fig.~\ref{fig:mean_field_regimes}d).

\section{Fixed point stability in the mean-field network}\label{app:DMFT_FP_stability}

Here we consider the full matrix of linear response functions (see below), to conclude that the only quantity that matters for the stability at the fixed point is $\tilde{G}(f)$.

Starting from the microscopic network equations (Eq.~\ref{eq:multi-d-micro}), we derive a set of differential equations, that we write in matrix form
\begin{equation}\label{eq:chi_starting_point_multi-d}
\left( \mathrm{I}_D\partial_\tau - \mathrm{A}\right) \chi_{ik}(\tau) = \sum_{j=1}^N J_{ij}\Delta_1 \chi_{jk}(\tau) + \delta_{ik}\mathrm{I}_D \delta(\tau) \quad ,
\end{equation}
where $\Delta_1=\delta^{\alpha 1}\delta^{\beta 1}$ is a matrix whose only nonzero element is $[\Delta_1]^{11}=1$.
$ \chi_{ik}(\tau)$ is a $D$ by $D$ matrix, whose component are defined as $ \chi_{ik}^{\alpha\beta}(\tau)=\frac{\delta x_i^\alpha(\tau)}{\delta h_k^\beta(0)}$, where $h_k^\beta$ is a small perturbation given to the variable $x_k^\beta$ at time $\tau=0$.
Notice that in deriving Eq.~\ref{eq:chi_starting_point_multi-d}, we have assumed stationarity and that $\phi'(0)=1$.
We now Fourier transform Eq.~\ref{eq:chi_starting_point_multi-d} and get
\begin{equation}
\left( 2\pi i f \, \mathrm{I}_D - \mathrm{A}\right) \tilde{\chi}_{ik}(f) = \sum_{j=1}^N J_{ij}\Delta_1 \tilde{\chi}_{jk}(f) + \delta_{ik}\mathrm{I}_D  \quad .
\end{equation}
Inverting the matrix $\left( 2\pi i f \, \mathrm{I}_D - \mathrm{A}\right)$ and recognizing the linear response function of the single unit $\tilde{\chi}_0(f)$, we obtain
\begin{equation}\label{eq:chi_mean_multi-d}
 \tilde{\chi}_{ik}(f) = \sum_{j=1}^N J_{ij}\tilde{\chi}_0(f)\Delta_1 \tilde{\chi}_{jk}(f) + \delta_{ik}\tilde{\chi}_0(f)  \quad ,
\end{equation}
where $\tilde{\chi}_0(f)$ is a $D$ by $D$ matrix whose elements are $\tilde{\chi}_0^{\alpha\beta}(f)$, defined in section \ref{sec:mean-field}.

Since in the mean-field approximation the mean of the linear response function is zero, we look for the second moments \cite{Kadmon15}.
We multiply every element of the matrix equation (Eq.~\ref{eq:chi_mean_multi-d}) by its complex conjugate and average over the quenched disorder.
We obtain
\begin{equation}
|\tilde{\chi}(f)|^2 = g^2 | \tilde{\chi}_0(f)\Delta_1 \tilde{\chi}(f)|^2 + \tilde{G}(f) \quad ,
\end{equation}
where the absolute value is intended element-wise.
Due to the structure of the matrix $\Delta_1$, we have that $|\tilde{\chi}_0(f)\Delta_1 \tilde{\chi}(f)|^2 = \tilde{G}(f) \Delta_1 |\tilde{\chi}(f)|^2$, as it can be verified simply by using the definition of $\Delta_1$.
Finally, we can solve for $|\tilde{\chi}(f)|^2$
\begin{equation}
|\tilde{\chi}(f)|^2 = \left( \mathrm{I}_D - g^2 \tilde{G}(f) \Delta_1 \right)^{-1} \left( \tilde{G}(f)  \right) \quad .
\end{equation}
Since the only nonzero eigenvalue of the matrix $\tilde{G}(f) \Delta_1 $ is $|\tilde{\chi}_0^{11}(f)|^2 $, the stability condition for the fixed point is given by
\begin{equation}
g^2 \max_f \tilde{G}(f) < 1 \quad .
\end{equation}

\section{Effect of nonlinearities on second-order statistics}\label{app:nonlinearities}
In this section, we provide some additional details on how to compute the effect of nonlinearities on the second order statistics (autocorrelation or power spectral density) of a Gaussian process.
We consider three cases of interest: polynomials, piecewise linear functions and arbitrary nonlinear functions.
To simplify our notation, we drop the superscript of and consider a generic Gaussian process $x^1$.

The effect of polynomial nonlinearities can be expressed in closed form in time domain.
This can be seen by considering again the infinite series expression (Eq.~\ref{eq:infinite_series_x1}), valid for stationary Gaussian processes $x$ 
\begin{equation}
\label{eq:cubic-map}
C_{\phi(x)}(\tau) = \sum_{n=0}^\infty \left(\left\langle \frac{d^n \phi}{dx^n}\right\rangle\right)^2 C_x^n(\tau) \quad ,
\end{equation}
where the angular brackets indicate the average over the statistics of $x$.
In the case in which $\phi$ is a polynomial of degree $p$, only the terms in the sum up to $p$ are nonzero.
As an example, we can compute the effect of a cubic approximation of the hyperbolic tangent, i.e. $\phi(x) \simeq \phi_3(x) := x - \frac{x^3}{3}$
\begin{equation}
C_{\phi_3(x)}(\tau) = \left(1+C_x^2(0) -2C_x(0)\right) C_x(\tau) + \frac{2}{3}C_x^3(\tau) \quad .
\end{equation}
As expected, the effect of the nonlinearity depends on $C_x(0)$ i.e. on the variance of $x$ itself.
Notice that the coefficient of the first term is compressive (i.e. smaller than one) only if $C_x(0)$ is smaller than one itself.
This type of behavior is expected since $\phi_3$ is unbounded.

Another interesting case are piecewise linear nonlinearities.
In this case, we use Price's theorem twice to get
\begin{equation}\label{eq:Price_twice}
\frac{\partial^2 C_{\phi(x)}(t)}{\partial (C_x(t))^2} = C_{\phi''(x)}(t) \quad .
\end{equation}
For a piecewise linear $\phi$, the second derivative $\phi''$ is a sum of Dirac's delta functions with variable coefficients.
More precisely, we consider
\begin{equation}
\begin{aligned}
\phi_{PL}(x) = & \Theta(x_1-x)c_{0}x \\
 + & \sum_{p=1}^{P-1}  \Theta(x-x_p)\Theta(x_{p+1}-x)c_p x_p + \Theta(x-x_P)c_P x \quad ,
\end{aligned}
\end{equation}
where $x_p$ are the points in which the first derivative is discontinuous, $c_p$ are some arbitrary coefficients and $\Theta(\cdot)$ is the Heaviside function.
The second derivative of $\phi_{PL}$ is given by
\begin{equation}
\phi_{PL}''(x) = \sum_{p=1}^{P}(c_p - c_{p-1})\delta(x-x_p) \quad .
\end{equation}
The delta functions allow us to compute the correlation function $C_{\phi_{PL}''}(t)$ explicitly
\begin{equation}\label{eq:cPhi_PL}
\begin{aligned}
C_{\phi_{PL}''}(t) =& \sum_{p,p'=1}^{P}\frac{(c_p - c_{p-1})(c_{p'} - c_{p'-1})}{2\pi C_x(0)\sqrt{1-\rho^2(t)}} \quad \times \\
\times \quad & \exp \left( -\frac{x_p^2 + x_{p'}^2 - 2\rho(t) x_p x_{p'}}{2C_x(0)(1-\rho^2(t) ) } \right)
\quad ,
\end{aligned}
\end{equation}
where we defined $\rho(t):= \frac{C_x(t)}{C_x(0)}$.
Inserting Eq.~\ref{eq:cPhi_PL} in Eq.~\ref{eq:Price_twice} and integrating twice with respect to $C_x(t)$ we get
\begin{equation}\label{eq:PWL_transform}
\begin{aligned}
C_{\phi_{PL}(x)}(t) = & f_{\phi}\left(0;C_x(0)\right) + f_{\phi'}\left(0;C_x(0)\right)C_x(t) \\
+ & \sum_{p,p'=1}^{P}\int_0^{C_x(t)}\int_0^{\sigma'}\frac{(c_p - c_{p-1})(c_{p'} - c_{p'-1})}{2\pi C_x(0)\sqrt{1-\frac{\sigma^2}{C_x^2(0)}} } \,  \times  \\
\times \, & \exp \left( -\frac{x_p^2 + x_{p'}^2 - 2\frac{\sigma}{C_x(0)} x_p x_{p'}}{2C_x(0)\left(1-\frac{\sigma^2}{C_x^2(0)} \right) } \right) d\sigma d\sigma' \, .
\end{aligned}
\end{equation}
In the case in which $\phi$ is an odd function, the term $ f_{\phi}\left(0;C_x(0)\right)$ is equal to zero. 
For the specific case of the piecewise linear approximation of the hyperbolic tangent considered in this paper, i.e.
\begin{equation}
\phi_{PL}(x) = \begin{cases}
-1 \quad \text{for} \quad x<-1 \\
x \quad \text{for} \quad -1<x<1\\
1 \quad \text{for} \quad x>1
\end{cases} \quad ,
\end{equation}
the expression in Eq.~\ref{eq:PWL_transform} reduces to
\begin{widetext}
\begin{align}\label{eq:PWL_transform_used}
C_{\phi_{PL}(x)}(t) = \text{Erf}^2\left(\frac{1}{\sqrt{2C_x(0)}}\right) C_x(t) +
\frac{2}{\pi C_x(0)}\int_0^{C_x(t)}\int_0^{\sigma'} \frac{1}{\sqrt{1-\frac{\sigma^2}{C_x^2(0)}}} \quad \times \nonumber \\
\times \quad \exp\left(-\frac{1}{C_x(0)\left(1-\frac{\sigma^2}{C_x^2(0)}\right)}\right)
\sinh\left(\frac{\sigma}{C_x^2(0)\left(1-\frac{\sigma^2}{C_x^2(0)}\right)}\right)
d\sigma d\sigma' \quad .
\end{align}
\end{widetext}
For the piecewise linear function, an alternative approach is based on the infinite series in Eq.~\ref{eq:infinite_series_x1}, which yields  \cite{Str67I,Kruscha16}:
\begin{equation}
  \label{eq:pl-inf-series}
  C_{\phi_{PL}(x)}(t)=\sigma^2\sum_{n=1}^\infty \left[F^{(n-1)}\left(\frac{1}{\sigma}\right)-F^{(n-1)}\left(\frac{-1}{\sigma}\right)\right]^2\frac{C_x^n(t)}{n!}
\end{equation}
with input variance $\sigma^2=C_x(0)$ and cumulative Gaussian distribution function $F(x)=\frac{1}{\sqrt{2\pi}}\int_{-\infty}^xe^{-y^2/2}\,dy$.
For the figures in this paper, we used the map in Eq.~\ref{eq:PWL_transform_used}.

For an arbitrary nonlinear function, we can use two methods.
The first method is a semi-analytical approach that relies on the integral form of the autocorrelation of the rate $C_{\phi(x)}(\tau)$ as a functional of the autocorrelation $C_{x}(\tau)$ of $x$ \cite{Schucker16a}
\begin{widetext}
\begin{equation}\label{eq:integral_form_Cphi}
C_{\phi(x)}(\tau) = \int\int \phi\left(\sqrt{C_x(0) - \frac{C_x^2(\tau)}{C_x(0)}}x  + \frac{C_x(\tau)}{\sqrt{C_x(0)}}z\right)\phi\left(\sqrt{C_x(0)}z\right)Dx Dz \quad ,
\end{equation}
\end{widetext}
where $Dx=e^{-x^2/2}dx$.
Notice that a slightly different version of this formula was already proposed in \cite{Sompolinsky88}.
Therefore, to obtain the effect of $\phi$ on the power spectral density, one should 1) inverse Fourier transform $S_x(f)$ to get $C_x(\tau)$ 2) apply Eq.(\ref{eq:integral_form_Cphi}), by computing the two integrals numerically 3) Fourier transform $C_{\phi(x)}(\tau)$ to get $S_{\phi(x)}(f)$.
Practically, this procedure requires the application of the fast Fourier transform algorithm and the numerical evaluation of two integrals.

The second method is fully numerical and it can be useful in cases in which the integrals in the first method are expensive to evaluate numerically.
This method consists in approximating the power spectral density $S_{\phi(x)}$ via Monte Carlo sampling.
More precisely, we sample multiple realizations in frequency domain of the Gaussian process with zero mean and power spectral density $S_x(f)$.
We then transform each sample to time domain and apply the nonlinearity $\phi(x)$ to each sample $x(t)$ individually.
Finally, we transform back to Fourier domain and get $S_{\phi(x)}$ by averaging.
Despite being computationally more expensive than the closed form expressions, this sampling method provides a solution of the mean-field theory for an arbitrary nonlinearity and it is computationally much cheaper than running the full microscopic simulation.
Moreover, this method can easily be extended to be used in the presence of a non-Gaussian sinusoidal input (cf. section \ref{sec:driven} and \cite{Rajan10}).

\end{document}